\documentclass[aps,onecolumn]{revtex4}
\usepackage{graphicx}
\usepackage{epsfig}
\usepackage{epstopdf}
\usepackage{amsfonts}
\usepackage{amssymb}
\usepackage{amsbsy}
\usepackage{amsmath}
\usepackage{mathrsfs}
\usepackage{latexsym}
\usepackage{natbib}
\usepackage{bm}
\usepackage{color}
\usepackage[a4paper, total={6.7in, 10.1in}]{geometry}

\usepackage{braket}
\usepackage{slashed}
\usepackage{pgfplots}
\numberwithin{equation}{section}

\usepackage{tikz}
\usetikzlibrary{shapes,arrows,shadows}

\begin{document}

\title{Characteristics of interaction between Gravitons and Photons}

\author{Athira B S}
\email{abs16rs013@iiserkol.ac.in}

\affiliation{ Center of Excellence in Space Sciences India,\\ 
Indian Institute of Science Education and Research Kolkata,\\
Mohanpur - 741 246, WB, India  }

\author{Susobhan Mandal}
\email{sm17rs045@iiserkol.ac.in}

\affiliation{ Department of Physical Sciences,\\ 
Indian Institute of Science Education and Research Kolkata,\\
Mohanpur - 741 246, WB, India }

\author{Subhashish Banerjee}
\email{subhashish@iitj.ac.in}

\affiliation{ Indian Institute of Technology Jodhpur, India }


\begin{abstract}
The direct detection of gravitational waves from binary mergers has been hailed as the discovery of the century. In the light of recent evidence on the existence of gravitational waves, it is now possible to know about the properties of matter under extreme conditions in compact astrophysical objects and different dynamical spacetimes. 
The foremost theme of the present article is to bring out the various features of the interaction between photons and gravitons that can be used in astrophysical observations. The effective action of interacting photons containing light-matter coupling and self-interaction term is constructed by eliminating the graviton degrees of freedom coupled to both matter and photons. It is shown that the equation of state of matter can be probed from the dynamics of light in this theory. The vacuum birefringence is also shown to be a generic property in this theory that arises from the non-linear nature of the self-interaction between gauge fields. Further, the non-local nature of quantum effective action with modified dispersion relation is also discussed in great detail. The above results also open an alternate way to infer the properties of gravitational waves without their direct measurement using the features of photon-graviton interaction.   
\end{abstract}

\maketitle

\section{Introduction}

The existence of gravitation waves is one of the most important features of Einstein's theory of general relativity (GR) and was first predicted by Einstein. The discovery of gravitational waves can be seen as a test to verify GR and also put constraints on alternate theories of gravity. Its discovery after a hiatus of almost a century from its theoretical prediction is because of the extreme sensitivities of the measurements involved. LIGO confirmed the first detection \cite{abbott2016observation, abbott2016gw151226, scientific2017gw170104, abbott2017gw170817, abbott2017gw170814, abbott2016binary} through laser optical interferometry.

Among the four fundamental interactions in nature, the electromagnetic and gravitational interactions are long-ranged and mediated by a spin-1 massless
photon and a spin-2 massless graviton, respectively. This property is useful and often used in order to probe different astrophysical observations. The success of LIGO in detecting gravitational waves motivated suggestions for utilizing optical measurement techniques that would incur lesser expense \cite{callister2017polarization, ford2019multi, Ni:2012eh, graham2013new, cahill2007optical, ali2019electromagnetic, nishizawa2014measuring, feng2019detect, oancea2019overview}. To make optical measurements, knowledge of the interaction of gravitational waves (GW) with light would be required. This question has been first attempted in \cite{weber1962interaction,skobelev1975graviton} where different scattering processes between gravitons and photons were studied \cite{bjerrum2015graviton, bartolo2018photon}. However, the problem with measuring such scattering amplitudes or cross-sections is that their numerical values are extremely small. Measurement of a single graviton is difficult with current technologies \cite{dyson2014graviton}. Nevertheless, it can capture useful physical information of spacetime which could be probed through weak measurement techniques, as will be shown in this article. The importance of studying the interaction between photons and gravitons has also been highlighted in \cite{baryshev2019solution, baryshev2017foundation}, in a different context.

In this article, we show how certain features of graviton-photon interactions can be accessed through optical measurements that indirectly confirm features of spacetime carried by gravitational waves. Throughout our discussion, we also comment on massive gravity theory which is an alternate theory of GR, motivated for solving the problem of Dark Energy and the current accelerated expansion of the Universe, among others. In massive gravity theory \cite{de2014massive, hinterbichler2012theoretical} (which are also ghost-free \cite{de2011resummation}), the infrared (IR) region of GR is modified by the addition of a mass term leading to gravitons becoming massive and spin-2.

The extremely small numerical values of scattering amplitudes of the graviton-photon interactions would suggest that a tool to amplify the signals involved would be very welcome. Such a scenario is facilitated by recent developments in the field of quantum optics, in particular, the weak measurement technique \cite{zhou2018photonic, zhou2012experimental, feizpour2011amplifying, dressel2018strengthening, audretsch2002quantum, smith2004continuous, lundeen2009experimental}. Weak measurements is the name coined to a measurement scenario in quantum mechanics, wherein the empirically measured value (called the {\it weak value}) of an observable can yield results beyond the eigenvalue spectrum of the measured observable. This has lead to a number of interesting developments including weak value amplification, useful for enhancing the sensitivity of specific detection schemes.

We briefly discuss the weak field limit of general relativity, followed by the Fierz-Pauli action of massive gravity and Stueckelberg's technique for restoring gauge symmetry to massive gravity action. This sets the scene for the construction of an effective action for interacting photons that takes into account the interactions between photons and gravitons. This is followed by some non-trivial features of on-shell equations obtained from the minimization of the effective action. A few scattering amplitudes are next computed between photonic states and it is shown that through weak measurement protocol these amplitudes can be amplified. Finally, one-loop quantum corrections are taken into account in order to write quantum effective action for the interacting photons, explicitly at the quadratic level. This leads to a modified dispersion relation for low-energy photons. Further, effective interacting vertices at the quantum level between photons are obtained. The quantum effective action for photons is shown to be non-local in nature, a generic property in this theory.

\section{Weak field limit of General Relativity}
\subsection{Free-field theory of massless gravitons}
The weak-field limit of the Einstein-Hilbert action in the presence of matter, considering the first-order correction to metric ($g=\eta+h$) is given by
\begin{equation}
\begin{split}
S_{EH} & =\int d^{4}x\mathcal{L}_{EH}+S_{M}\\
\mathcal{L}_{EH} & =-h_{ \ \sigma}^{\mu}\partial^{\sigma}\partial^{\nu}h_{\mu\nu}+h\partial^{\mu}\partial^{\nu}h_{\mu\nu}+\frac{1}{2}h_{\mu\nu}\Box h^{\mu\nu}-\frac{1}{2}h\Box h\\
S_{M}(\eta+h) & =S_{M}|_{\eta}+\int d^{4}x\left(\frac{1}{\sqrt{-g}}\frac{\delta S_{M}}{\delta g^{\mu\nu}}\right)\Big|_{\eta}h^{\mu\nu},
\end{split}
\end{equation}
where the invariance of the action under the transformation $h_{\mu\nu}\rightarrow h_{\mu\nu}+\partial_{\mu}\xi_{\nu}+\partial_{\nu}\xi_{\mu}$ can be checked explicitly. The Lagrangian can be recast as
\begin{equation}
\begin{split}
\mathcal{L}_{EH} & =\frac{1}{2}h_{\mu\nu}\mathcal{O}^{\mu\nu\rho\sigma}h_{\rho\sigma}\\
\mathcal{O}^{\mu\nu\rho\sigma} & =\left(\frac{1}{2}\eta^{\mu\rho}\eta^{\nu\sigma}+\frac{1}{2}\eta^{\mu\sigma}\eta^{\nu\rho}-\eta^{\mu\nu}\eta^{\rho\sigma}\right)\Box+\eta^{\mu\nu}\partial^{\rho}\partial^{\sigma}+\eta^{\rho\sigma}\partial^{\mu}\partial^{\nu}\\
 & -\frac{1}{2}(\eta^{\nu\rho}\partial^{\mu}\partial^{\sigma}+\eta^{\nu\sigma}\partial^{\mu}\partial^{\rho}+\eta^{\mu\rho}\partial^{\nu}\partial^{\sigma}+\eta^{\mu\sigma}\partial^{\nu}\partial^{\rho}).
\end{split}
\end{equation}
The corresponding equation of motion becomes
\begin{equation}
\mathcal{O}^{\mu\nu\rho\sigma}h_{\rho\sigma}=8\pi GT^{\mu\nu},
\end{equation}
which implies $\partial_{\mu}\mathcal{O}^{\mu\nu\rho\sigma}=0$. Therefore, in momentum space the operator $\mathcal{O}^{\mu\nu\rho\sigma}$ is not invertible, and hence, we can not have a corresponding Green's function. That is expected because this is a gauge-invariant theory and the gauge has not yet been fixed which is done in the next section. 
\subsection{Gauge-fixing}
We introduce the following gauge fixing term, the de Donder gauge \cite{hinterbichler2012theoretical},
\begin{equation}
\mathcal{L}_{GF}=-\frac{1}{\alpha}(\partial_{\rho}h_{ \ \mu}^{\rho}-\frac{1}{2}\partial_{\mu}h)(\partial_{\sigma}h^{\mu\sigma}-\frac{1}{2}\partial^{\mu}h),
\end{equation}
where $\alpha$ is known as gauge parameter. With this new term, the weak field lagrangian density becomes
\begin{equation}
\begin{split}
\bar{\mathcal{L}}_{EH} & =\mathcal{L}_{EH}+\mathcal{L}_{GF}, \ \mathcal{L}_{GF}=\frac{1}{2}h_{\mu\nu}\mathcal{O}_{GF}^{\mu\nu\rho\sigma}h_{\rho\sigma},\\
\mathcal{O}_{GF}^{\mu\nu\rho\sigma} & =\frac{1}{\alpha}\left(2\eta^{\mu\rho}\partial^{\nu}\partial^{\sigma}-\eta^{\rho\sigma}\partial^{\mu}\partial^{\nu}-\eta^{\mu\nu}\partial^{\rho}\partial^{\sigma}+\frac{1}{2}\eta^{\mu\nu}\eta^{\rho\sigma}\Box\right),\\
\implies\bar{\mathcal{L}}_{EH} & =\frac{1}{2}h_{\mu\nu}(\mathcal{O}+\mathcal{O}_{GF})^{\mu\nu\rho\sigma}h_{\rho\sigma},
\end{split}
\end{equation}
where
\begin{equation}
\begin{split}
\tilde{\mathcal{O}} & =\mathcal{O}+\mathcal{O}_{GF}\\
\tilde{\mathcal{O}}^{\mu\nu\rho\sigma} & =\left(\frac{1}{2}\eta^{\mu\rho}\eta^{\nu\sigma}+\frac{1}{2}\eta^{\mu\sigma}\eta^{\nu\rho}-\left(1-\frac{1}{2\alpha}\right)\eta^{\mu\nu}\eta^{\rho\sigma}\right)\Box+\left(1-\frac{1}{\alpha}\right)(\eta^{\mu\nu}\partial^{\rho}\partial^{\sigma}+\eta^{\rho\sigma}\partial^{\mu}\partial^{\nu})\\
 & +\frac{1}{2}\left(\frac{1}{\alpha}-1\right)(\eta^{\nu\rho}\partial^{\mu}\partial^{\sigma}+\eta^{\nu\sigma}\partial^{\mu}\partial^{\rho}+\eta^{\mu\rho}\partial^{\nu}\partial^{\sigma}+\eta^{\mu\sigma}\partial^{\nu}\partial^{\rho}).
\end{split}
\end{equation}
Therefore, after choosing the Feynman gauge $\alpha=1$, in momentum space, the Green's function takes the following form
\begin{equation}
\Pi_{GR,\mu\nu\rho\sigma}=-\frac{i}{2k^{2}}(\eta_{\mu\rho}\eta_{\nu\sigma}+\eta_{\mu\sigma}\eta_{\nu\rho}-\eta_{\mu\nu}\eta_{\rho\sigma}).
\end{equation}
From now onwards $\tilde{\mathcal{O}}$ will be denoted as $\mathcal{O}$ for sake of convenience. Gauge fixing could also have been achieved using the Faddeev-Popov Ghost \cite{faddeev2010faddeev, eichhorn2013faddeev} method in the path integral formalism.

\section{Massive Gravity}
The unique action that describes a massive spin-2 particle in flat spacetime in which field is described by a symmetric rank-2 tensor is
\begin{equation}\label{action1}
S=\int d^{D}x\Big[-\frac{1}{2}\partial_{\lambda}h_{\mu\nu}\partial^{\lambda}h^{\mu\nu}
+\partial_{\mu}h_{\nu\lambda}\partial^{\nu}h^{\mu\lambda}-\partial_{\mu}h^{\mu\nu}\partial_{\nu}h+\frac{1}{2}\partial_{\lambda}h\partial^{\lambda}h-\frac{1}{2}m^{2}(h_{\mu\nu}h^{\mu\nu}-h^{2})\Big],
\end{equation}
known as the Fierz-Pauli action \cite{blasi2017massive, folkerts2013massive, de2010generalization}. Note that when $m=0$ this becomes the linearized Einstein-Hilbert action, invariant under the following gauge transformation
\begin{equation}
\delta h_{\mu\nu}=\partial_{\mu}\xi_{\nu}+\partial_{\nu}\xi_{\mu}.
\end{equation}
The above action is not gauge invariant, but will be made so by using Stueckelberg's trick \cite{noller2014interacting, ruegg2004stueckelberg, kors2005aspects, hinterbichler2012theoretical} to massive gravity action in order to restore gauge symmetry. Introducing Stueckelberg's auxiliary fields $V^{\mu}, \phi$ with specific gauge fixing terms
\begin{equation}
\begin{split}
S_{GF1} & =-\int d^{4}x\left(\partial^{\nu}h_{\mu\nu}-\frac{1}{2}\partial_{\mu}h+mV_{\mu}\right)^{2}, \ S_{GF2}=-\int d^{4}x\left(\partial_{\mu}V^{\mu}+m\left(\frac{1}{2}h+3\phi\right)\right)^{2},
\end{split}
\end{equation}
the Fierz-Pauli action with a source can be written as (a detailed derivation is provided in the Appendix)
\begin{equation}\label{action2}
\begin{split}
S+S_{GF1}+S_{GF2} & =\int d^{4}x\Big[\frac{1}{2}h_{\mu\nu}(\Box-m^{2})h^{\mu\nu}-\frac{1}{4}h(\Box-m^{2})h+V_{\mu}(\Box-m^{2})V^{\mu}\\
+3\phi(\Box-m^{2})\phi & +\kappa h_{\mu\nu}T^{\mu\nu}+\kappa\phi T-\frac{2}{m}\kappa V_{\mu}\partial_{\nu}T^{\mu\nu}+\frac{2}{m^{2}}\kappa\phi\partial_{\mu}\partial_{\nu}T^{\mu\nu}\Big].
\end{split}
\end{equation}
For transverse and traceless energy-momentum tensor, the last three terms of the above action vanish. This makes vector and scalar degrees of freedom completely decoupled from interaction with matter. The propagators of $h_{\mu\nu}, \ V_{\mu}, \ \phi$ in momentum space are now 
\begin{equation}
\begin{split}
-\frac{i}{p^{2}+m^{2}}\frac{1}{2} & (\eta_{\mu\alpha}\eta_{\nu\beta}+\eta_{\mu\beta}\eta_{\nu\alpha}-\eta_{\mu\nu}\eta_{\alpha\beta}), \ -\frac{i}{2}\frac{\eta_{\mu\nu}}{p^{2}+m^{2}}, \ -\frac{i}{6(p^{2}+m^{2})},
\end{split}
\end{equation}
respectively. They all behave as $\frac{1}{p^{2}}$ for large momenta, implying that standard power counting arguments are applicable.
 
\section{Photon-Graviton interaction}
\subsection{Introduction}
Consider a source whose stress-energy tensor $T_{\mu\nu}^{(c)}$ produces gravitational waves (GW) that travel through spacetime to asymptotically flat spacetime and interacts with a medium of photons. Our expectation is that the interaction between photons and gravitons captures the properties of the original source of GW.

The action for such a system would be
\begin{equation}
\begin{split}
S & =S_{m=0}^{(spin-2)}+S_{GF}+S_{\text{photon}}=\int d^{4}x\Big[\frac{1}{2}h_{\mu\nu}\mathcal{O}^{\mu\nu,\alpha\beta}h_{\alpha\beta}+\kappa h_{\mu\nu}T^{(c)\mu\nu}-\frac{1}{4}F_{\mu\nu}F^{\mu\nu}+\kappa h_{\mu\nu}T^{(s=1)\mu\nu}\Big],
\end{split}
\end{equation}
where $\kappa=\sqrt{\frac{8\pi G\hbar}{c^{4}}}$, $F_{\mu\nu}=\partial_{\mu}A_{\nu}-\partial_{\nu}A_{\mu}$ and $T^{(s=1)\mu\nu}$ is the stress-energy tensor of photons.
Therefore, the generating functional can be written as
\begin{equation}
\begin{split}
\mathcal{Z} & [J^{\mu\nu}=0]=\int\mathcal{D}h_{\mu\nu}\mathcal{D}A_{\mu}e^{i\int d^{4}x\Big[\frac{1}{2}h_{\mu\nu}\mathcal{O}^{\mu\nu,\alpha\beta}h_{\alpha\beta}+\kappa h_{\mu\nu}T^{(c)\mu\nu}-\frac{1}{4}F_{\mu\nu}F^{\mu\nu}+\kappa h_{\mu\nu}T^{(s=1)\mu\nu}\Big]}\\
=\mathcal{N} & e^{\frac{i}{2}\kappa^{2}\int d^{4}x T^{(c)\mu\nu}\mathcal{D}_{\mu\nu\alpha\beta}T^{(c)\alpha\beta}}\int\mathcal{D}A_{\mu}e^{i\int d^{4}x\Big[-\frac{1}{4}F_{\mu\nu}F^{\mu\nu}+\kappa^{2}T^{(c)\mu\nu}\mathcal{D}_{\mu\nu\alpha\beta}T^{(s=1)\alpha\beta}+\frac{\kappa^{2}}{2}T^{(s=1)\mu\nu}\mathcal{D}_{\mu\nu\alpha\beta}T^{(s=1)\alpha\beta}\Big]}.
\end{split}
\end{equation}
By integrating out the graviton degrees of freedom, we obtain the following effective action for the photon degrees of medium
\begin{equation}\label{eff1}
S_{eff}^{(s=1)}=\int d^{4}x\Big[-\frac{1}{4}F_{\mu\nu}F^{\mu\nu}+\kappa^{2}T^{(c)\mu\nu}\mathcal{D}_{\mu\nu\alpha\beta}T^{(s=1)\alpha\beta}+\frac{\kappa^{2}}{2}T^{(s=1)\mu\nu}\mathcal{D}_{\mu\nu\alpha\beta}T^{(s=1)\alpha\beta}\Big].
\end{equation}
As can be seen from the above equation, the third piece is purely an interacting term, taking into account the effective interaction between photons. For the time being, the interaction term is neglected; which is justified for weak gauge fields, the action involves only quadratic or free part and reduces to
\begin{equation}\label{eff2}
S_{eff}^{(1)}=\int d^{4}x\Big[-\frac{1}{4}F_{\mu\nu}F^{\mu\nu}+\kappa^{2}T^{(c)\mu\nu}\mathcal{D}_{\mu\nu\alpha\beta}T^{(s=1)\alpha\beta}\Big],
\end{equation}
where (for massless gravitons)
\begin{equation}
\begin{split}
T^{(s=1)\mu\nu} & =\eta_{\alpha\beta}F^{\alpha\mu}F^{\beta\nu}-\frac{1}{4}\eta^{\mu\nu}F_{\rho\sigma}F^{\rho\sigma}, \ \mathcal{D}_{\mu\nu\alpha\beta}=\frac{1}{2\Box}(\eta_{\mu\alpha}\eta_{\nu\beta}+\eta_{\mu\beta}\eta_{\nu\alpha}-\eta_{\mu\nu}\eta_{\alpha\beta}).
\end{split}
\end{equation}
Therefore,
\begin{equation}
\begin{split}
T^{(c)\mu\nu}\mathcal{D}_{\mu\nu\alpha\beta}T^{(s=1)\alpha\beta} & =T^{(c)\mu\nu}\frac{1}{2\Box}(T_{\mu\nu}^{(s=1)}+T_{\mu\nu}^{(s=1)}-\underbrace{T^{(s=1)}}_{=0}\eta_{\mu\nu})\\
=\left(\frac{1}{\Box}T^{(c)\mu\nu}\right)\Big[\eta^{\alpha\beta} & (\partial_{\alpha}A_{\mu}\partial_{\beta}A_{\nu}-\partial_{\alpha}A_{\mu}\partial_{\nu}A_{\beta}-\partial_{\mu}A_{\alpha}\partial_{\beta}A_{\nu}+\partial_{\mu}A_{\alpha}\partial_{\nu}A_{\beta})-\frac{1}{4}\eta_{\mu\nu}F_{\rho\sigma}F^{\rho\sigma}\Big].
\end{split}
\end{equation}
where the last line follows from integration by parts.
\subsection{Equations of motion}
The Lagrangian density in (\ref{eff2}) can be expressed as
\begin{equation}
\begin{split}
\mathcal{L} & =-\frac{1}{2}(\partial_{\mu}A_{\nu}\partial^{\mu}A^{\nu}-\partial_{\nu}A_{\mu}\partial^{\mu}A^{\nu})+\kappa^{2}\left(\frac{1}{\Box}T^{(c)\mu\nu}\right) \Big[\eta^{\alpha\beta}(\partial_{\alpha}A_{\mu}\partial_{\beta}A_{\nu}-\partial_{\alpha}A_{\mu}\partial_{\nu}A_{\beta}-\partial_{\mu}A_{\alpha}\partial_{\beta}A_{\nu}+\partial_{\mu}A_{\alpha}\partial_{\nu}A_{\beta})\Big]\\
 & -\kappa^{2}\left(\frac{1}{\Box}T^{(c)}\right)\frac{1}{4}F_{\rho\sigma}F^{\rho\sigma}.
\end{split}
\end{equation}
Therefore, the corresponding equations of motion are given by
\begin{equation}\label{eqn.1}
\begin{split}
-\left(1+\kappa^{2}\frac{1}{\Box}T^{(c)}\right) & \partial_{\rho}F^{\rho\sigma}-\kappa^{2}\frac{1}{\Box}\partial_{\rho}T^{(c)}F^{\rho\sigma}+2\kappa^{2}\Big[\frac{1}{\Box}\partial_{\rho}T^{(c)\sigma\nu}\partial^{\rho}A_{\nu}+\frac{1}{\Box}T^{(c)\sigma\nu}\Box A_{\nu}\\
-\frac{1}{\Box}\partial_{\rho}T^{(c)\sigma\nu}\partial_{\nu}A^{\rho} & -\frac{1}{\Box}T^{(c)\sigma\nu}\Box\omega_{\nu\rho}A^{\rho}-\frac{1}{\Box}T^{(c)\mu\rho}\Box\omega_{\rho}^{\sigma}A_{\mu}+\frac{1}{\Box}T^{(c)\rho\nu}\Box\omega_{\rho\nu}A^{\sigma}\Big]=0,
\end{split}
\end{equation}
where $\omega_{\mu\nu}=\frac{\partial_{\mu}\partial_{\nu}}{\Box}$, acts as a projection operator along longitudinal polarization of gauge fields. The above equation can be expressed as
\begin{equation}\label{longitudinal pol}
\begin{split}
-\left(1+\kappa^{2}\frac{1}{\Box}T^{(c)}\right) & \partial_{\rho}F^{\rho\sigma}-\kappa^{2}\frac{1}{\Box}\partial_{\rho}T^{(c)}F^{\rho\sigma}+2\kappa^{2}\Big[\frac{1}{\Box}\partial_{\rho}T^{(c)\sigma\nu}F_{ \ \nu}^{\rho}+\frac{1}{\Box}T^{(c)\sigma\nu}\Box[\eta_{\nu\rho}-\omega_{\nu\rho}]A^{\rho}\\
 & +\frac{1}{\Box}T^{(c)\rho\nu}\Box[\omega_{\rho\nu}A^{\sigma}-\omega_{\rho}^{ \ \sigma}A_{\nu}]\Big]=0,
\end{split}
\end{equation}
which shows the absence of longitudinal degree of gauge field in the equation of motion. This can be checked by decomposing the gauge field as $A_{\mu}=A_{\mu}^{T}+\partial_{\mu}\chi$ where $A_{\mu}^{T}$ is the transverse component of the gauge field satisfying $\partial^{\mu}A_{\mu}^{T}=0$, and $\chi$ is the longitudinal component of the gauge field. Hence, the longitudinal degree of the gauge field is non-dynamical and it does not play any role in the dynamics of this theory.

\subsection{Features of equations of motion}
The Sun of our solar system can be considered as a standard candle that acts as a source of a medium of photons. The change in the polarization state of the light emitted from the Sun carries a signature of the GW interacting with the solar photons ignoring the other light-matter interactions. In this section and later (in section \ref{quantum correction}), we show that the photons interacting with gravitons are massive in nature. As a consequence, under a general gauge transformation, the transversality condition is not satisfied by the polarization of the light. Further, as shown below, the helicity of light is not conserved in a scattering process between the photons coupled with GW in general. This principle can be used to detect GW using the polarization measurement of the light emitted from the Sun before and after GW passes by. Here, we want to emphasize that the equations of motion of photons interacting with GW carry the information of different properties of GW generating sources that are captured by the stress-energy tensor of the sources, for example, for binary mergers one can get information such as charge, spin, angular momentum and mass of these compact objects.

Another important feature that would help us to put constraints on the graviton mass and IR domain of GR is that if we consider massive gravitons \cite{hinterbichler2012theoretical}, the equations of motion simply turn into
\begin{equation}\label{eqn.2}
\begin{split}
-\left(1+\kappa^{2}\frac{1}{\Box+m^{2}}T^{(c)}\right) & \partial_{\rho}F^{\rho\sigma}-\kappa^{2}\frac{1}{\Box-m^{2}}\partial_{\rho}T^{(c)}F^{\rho\sigma}+2\kappa^{2}\Big[\frac{1}{\Box-m^{2}}\partial_{\rho}T^{(c)\sigma\nu}\partial^{\rho}A_{\nu}+\frac{1}{\Box-m^{2}}T^{(c)\sigma\nu}\Box A_{\nu}\\
-\frac{1}{\Box-m^{2}}\partial_{\rho}T^{(c)\sigma\nu}\partial_{\nu}A^{\rho} & -\frac{1}{\Box-m^{2}}T^{(c)\sigma\nu}\Box\omega_{\nu\rho}A^{\rho}-\frac{1}{\Box-m^{2}}T^{(c)\mu\rho}\Box\omega_{\rho}^{\sigma}A_{\mu}+\frac{1}{\Box-m^{2}}T^{(c)\rho\nu}\Box\omega_{\rho\nu}A^{\sigma}\Big]=0,
\end{split}
\end{equation}
where $m$ is the mass of gravitons which follows from the action (\ref{action1}). For photons, the stress-energy tensor satisfies the following two important conditions
\begin{equation}
\partial_{\mu}T^{(s=1)\mu\nu}=0, \ \ T^{(s=1)}=0,
\end{equation}
which would kill the last three terms in the action (\ref{action2}). This suggests that in the presence of photons, vector and scalar degrees of freedom do not couple with photon degrees of freedom. Hence, they can be essentially treated as free-fields separately. It follows that these degrees of freedom can be integrated out without having any net effect on the effective action of photons, obtained earlier. Therefore, matching the data with (\ref{eqn.2}), it would be possible to put a constraint on the mass of gravitons in a similar manner to the constraint put by LIGO and others \cite{perkins2019probing, rana2018bounds, will2018solar, malsawmtluangi2017graviton, zakharov2018different, zakharov2017graviton, desai2019recent, gupta2018limit}.

Consider a compact object, comprising an ideal fluid, emitting gravitational waves. Its stress-energy tensor is 
\begin{equation}
T^{(c)\mu\nu}=(P+\rho)u^{\mu}u^{\nu}+P\eta^{\mu\nu},
\end{equation}
where $P, \ \rho$ are the pressure and the energy density of the matter inside the compact object. This kind of matter could be considered for other gravitational wave sources, though the equation of states may be different. $u^{\mu}$, a time-like unit vector is the velocity of fluid \textit{w.r.t} the observer. In this case, we obtain the following expression for the term in the action that describes the interaction between photons and a classical source
\begin{equation}
T^{(c)\mu\nu}\frac{1}{\Box}T_{\mu\nu}^{(s=1)}=\frac{1}{4}F_{\mu\nu}F^{\mu\nu}\frac{1}{\Box}(P+\rho)+F_{\alpha}^{ \ \mu}F^{\alpha\nu}\frac{1}{\Box}[(P+\rho)u_{\mu}u_{\nu}].
\end{equation}
Hence, the action in (\ref{eff2}) becomes the following
\begin{equation}\label{new action}
\begin{split}
S^{(1)} & =\int d^{4}x\Big[-\frac{1}{4}F_{\mu\nu}F^{\mu\nu}\left(1-\frac{\kappa^{2}}{\Box}(P+\rho)\right)+F_{\alpha}^{ \ \mu}F^{\alpha\nu}\frac{\kappa^{2}}{\Box}[(P+\rho)u_{\mu}u_{\nu}]\Big]\\
 & =\int d^{4}x\Big[-\frac{1}{4}F_{\mu\nu}(x)F^{\mu\nu}(x)[1+\Delta(x)]+F_{\alpha}^{ \ \mu}F^{\alpha\nu}\Delta_{\mu\nu}(x)\Big],
\end{split}
\end{equation}
where $\Delta_{\mu\nu}=\frac{\kappa^{2}}{\Box}[(P+\rho)u_{\mu}u_{\nu}]$ and $\Delta=\eta^{\mu\nu}\Delta_{\mu\nu}$. The equations of motion for the photon in this case are
\begin{equation}\label{new equation}
\begin{split}
-\partial_{\mu}[F^{\mu\nu}(x)(1+\Delta(x))] & +4\partial_{\mu}[F^{\mu\rho}(x)\Delta_{\rho}^{ \ \nu}(x)]=0\\
\implies -\partial_{\mu}F^{\mu\nu}(x)-F^{\mu\nu}(x)\partial_{\mu}\log[1+\Delta(x)] & +4\frac{\Delta_{\rho}^{ \ \nu}(x)}{1+\Delta(x)}\partial_{\mu}F^{\mu\rho}(x)+\frac{4}{1+\Delta(x)}F^{\mu\rho}(x)\partial_{\mu}\Delta_{\rho}^{ \ \nu}(x)=0.
\end{split}
\end{equation}
Since the action is gauge-invariant, hence, we fix a gauge $\partial_{\mu}A^{\mu}=0$. As a result of that, the above equation becomes
\begin{equation}
-\Box A^{\nu}(x)-F^{\mu\nu}(x)\partial_{\mu}\Delta^{(1)}(x)+4\Delta_{\rho}^{(2)\nu}(x)\partial_{\mu}F^{\mu\rho}(x)+4F^{\mu\rho}(x)\Delta_{\mu\rho}^{(3)\nu}(x)=0,
\end{equation}
where $\Delta^{(1)}(x)=\log(1+\Delta(x)), \ \Delta_{\rho}^{(2)\nu}(x)=\frac{\Delta_{\rho}^{ \ \nu}(x)}{1+\Delta(x)}, \ \Delta_{\mu\rho}^{(3)\nu}(x)=\frac{1}{1+\Delta(x)}\partial_{\mu}\Delta_{\rho}^{ \ \nu}(x)$. Expressing the gauge fields on the 4-momentum basis, the above equation can be expressed as
\begin{equation}\label{dispersion0}
\begin{split}
k^{2}A^{\nu}(k) & -\int\frac{d^{4}l}{(2\pi)^{4}}\Big[iF^{\mu\nu}(k-l)l_{\mu}\Delta^{(1)}(l)+4\Delta_{\rho}^{(2)\nu}(l)(k-l)^{2}A^{\rho}(k-l)-4F^{\mu\rho}(k-l)\Delta_{\mu\rho}^{(3)\nu}(l)\Big]=0\\
il_{\mu}F^{\mu\nu} & (k-l)=-[(k-l)_{\mu}l^{\mu} A^{\nu}(k-l)-(k-l)^{\nu}l_{\mu}A^{\mu}(k-l)]\\
F^{\mu\rho}(k-l) & \Delta_{\mu\rho}^{(3)\nu}(l)=i\Delta_{\mu\rho}^{(3)\nu}(l)[(k-l)^{\mu}A^{\rho}(k-l)-(k-l)^{\rho}A^{\mu}(k-l)],
\end{split}
\end{equation}
a non-local integral equation in momentum space. Further, the above equation clearly shows that the dynamics of a component of the gauge field depends on its other two components in general. The above on-shell relation clearly suggests that $k^{2}=0$ is not the dispersion of photons due to the presence of a non-local integral term. Under a gauge transformation $A_{\mu}\rightarrow A_{\mu}+\partial_{\mu}\Psi$, the polarization changes as $\epsilon_{\mu}(k)\rightarrow\epsilon'_{\mu}(k)=\epsilon_{\mu}(k)+ik_{\mu}\Psi(k)$. Since $k^{2}=0$ is not the dispersion or on-shell relation, under an arbitrary gauge transformation transversality condition is not maintained as $k^{\mu}\epsilon_{\mu}(k)\neq k^{\mu}\epsilon'_{\mu}(k)$, unlike free Maxwell field theory without gravitons. However, it is always possible to choose a gauge in which the polarization satisfies the transversality condition as the longitudinal polarization is not a dynamical component, shown in (\ref{longitudinal pol}). Since $k^{2}=0$ is not the dispersion of the photons interacting with GW in this case, photons have three transverse polarizations in this case, unlike the Maxwell field theory. This is also shown in section \ref{quantum correction}. The dimensionless functions and tensorial quantities defined above depends on the functions $P(x), \ \rho(x)$. Moreover, the non-locality of the above equation shows that the amplitude of a mode at on momentum $k$ depends on all the momentum modes. The non-locality property of the equations of motion originates from the source functions $P(x), \ \rho(x)$. From the equation (\ref{new equation}), we can also write the following first-order coupled partial differential equation 
\begin{equation}
-F^{\mu\nu}(x)(1+\Delta(x))+4F^{\mu\rho}(x)\Delta_{\rho}^{ \ \nu}(x)=\left(\eta^{\mu\nu}-\frac{\partial^{\mu}\partial^{\nu}}{\Box}\right)\mathcal{K}(x),
\end{equation}
with the unknown function $\mathcal{K}(x)$. However, in $2+1$-dimension, in principle, there could be another term in \textit{r.h.s} of the above equation which is of the form $\epsilon^{\mu\nu\lambda}\partial_{\lambda}\mathcal{W}(x)$.

Another point, we want to emphasize here is, the sign or rather the factor in front of the first kinetic term in (\ref{new action}) can be expressed as
\begin{equation}
\begin{split}
\left(1-\frac{\kappa^{2}}{\Box+i\epsilon}[P(x)+\rho(x)]\right) & =\Big[1+i\kappa^{2}\int\frac{d^{3}k}{(2\pi)^{3}}\frac{1}{2|\vec{k}|}[P(k)+\rho(k)]_{k^{0}=|\vec{k}|}e^{i\vec{k}.\vec{x}-i|\vec{k}|t}\\
 & -i\kappa^{2}\int\frac{d^{3}k}{(2\pi)^{3}}\frac{1}{2|\vec{k}|}[P(k)+\rho(k)]_{k^{0}=-|\vec{k}|}e^{i\vec{k}.\vec{x}+i|\vec{k}|t}\Big].
\end{split}
\end{equation}
Considering the isotropic pressure, i.e., $[P(k)+\rho(k)]_{k^{0}=|\vec{k}|}=[P(k)+\rho(k)]_{k^{0}=-|\vec{k}|}$ that depends only on $|\vec{k}|$ then, the above equation can be re-expressed as
\begin{equation}\label{new2}
\begin{split}
\left(1-\frac{\kappa^{2}}{\Box+i\epsilon}[P(x)+\rho(x)]\right) & =\Big[1+\frac{\kappa^{2}}{r}\int_{0}^{\infty}\frac{dk}{2\pi^{2}}[P(k)+\rho(k)]_{k^{0}=|\vec{k}|}\sin(kr)\sin(kt)\Big].
\end{split}
\end{equation}
It can be seen that as $r\rightarrow\infty$, the above term in the parenthesis becomes one. On the other hand, for $r\rightarrow0$, the above expression reduces to
\begin{equation}
\begin{split}
\left(1-\frac{\kappa^{2}}{\Box+i\epsilon}[P(x)+\rho(x)]\right) & \approx\Big[1+\kappa^{2}\int_{0}^{\infty}\frac{dk}{2\pi^{2}}[P(k)+\rho(k)]_{k^{0}=|\vec{k}|}k\sin(kt)\Big]\\
 & =\Big[1-\kappa^{2}\frac{d}{dt}\int_{0}^{\infty}\frac{dk}{2\pi^{2}}[P(k)+\rho(k)]_{k^{0}=|\vec{k}|}\cos(kt)\Big].
\end{split}
\end{equation}
This factor essentially affects the vacuum permeability which follows from the structure of Lagrangian density in QED. For the massive gravity theory, the equation (\ref{new2}) becomes
\begin{equation}
\begin{split}
\left(1-\frac{\kappa^{2}}{\Box-m^{2}+i\epsilon}[P(x)+\rho(x)]\right) & =\Big[1+\frac{\kappa^{2}}{r}\int_{0}^{\infty}\frac{dk}{2\pi^{2}}\frac{k}{\omega_{k}}[P(k)+\rho(k)]_{k^{0}=|\vec{k}|}\sin(kr)\sin(\omega_{k}t)\Big],
\end{split}
\end{equation}
where $\omega_{k}=\sqrt{k^{2}+m^{2}}$ and $m$ is the mass of gravitons. The above expression can also be expressed as
\begin{equation}
\begin{split}
\left(1-\frac{\kappa^{2}}{\Box-m^{2}+i\epsilon}[P(x)+\rho(x)]\right) & =\Big[1-\frac{\kappa^{2}}{r}\int_{0}^{\infty}\frac{dk}{2\pi^{2}}\omega_{k}\frac{d}{dk}\left([P(k)+\rho(k)]_{k^{0}=|\vec{k}|}\sin(kr)\sin(\omega_{k}t)\right)\Big].
\end{split}
\end{equation}
Hence, the above factor also depends on $\frac{d}{dk}[P(k)+\rho(k)]=\frac{d\rho(k)}{dk}\Big[1+\frac{dP(k)}{d\rho(k)}\Big]$.

\subsection{Source free gravitons interact with photons and birefringence}
If we now consider the source free gravitons interacting with photons then, $T^{(c)\mu\nu}=0$ in (\ref{eff1}), leading to
\begin{equation}
S_{eff}^{(s=1)}=\int d^{4}x\Big[-\frac{1}{4}F_{\mu\nu}F^{\mu\nu}+\frac{\kappa^{2}}{2}T^{(s=1)\mu\nu}\mathcal{D}_{\mu\nu\alpha\beta}T^{(s=1)\alpha\beta}\Big]=\int d^{4}x\Big[-\frac{1}{4}F_{\mu\nu}F^{\mu\nu}+\frac{\kappa^{2}}{2}T^{(s=1)\mu\nu}\frac{1}{\Box}T_{\mu\nu}^{(s=1)}\Big].
\end{equation}
and the equations of motion become
\begin{equation}
\begin{split}
-\partial_{\mu}F^{\mu\nu} & +\frac{\kappa^{2}}{2}\partial_{\mu}\Bigg[\left(\frac{\partial}{\partial(\partial_{\mu}A_{\nu})}(F_{ \ \sigma}^{\kappa}F^{\delta\sigma})\right)\frac{1}{\Box}(F_{\kappa}^{ \ \beta}F_{\delta\beta})\Bigg]+\frac{\kappa^{2}}{2}\partial_{\mu}\Bigg[F_{ \ \sigma}^{\kappa}F^{\delta\sigma}\frac{1}{\Box}\left(\frac{\partial}{\partial(\partial_{\mu}A_{\nu})}(F_{\kappa}^{ \ \beta}F_{\delta\beta})\right)\Bigg]\\
 & -\frac{\kappa^{2}}{4}\partial_{\mu}\Big[F^{\mu\nu}\frac{1}{\Box}(F_{\alpha\beta}F^{\alpha\beta})\Big]-\frac{\kappa^{2}}{4}\partial_{\mu}\Big[F_{\alpha\beta}F^{\alpha\beta}\frac{1}{\Box}F^{\mu\nu}\Big]=0.
\end{split}
\end{equation}
This can also be expressed as
\begin{equation}\label{source}
\partial_{\mu}F^{\mu\nu}=(\partial_{\mu}\mathcal{S}^{\mu\nu})\frac{\kappa^{2}}{2}\equiv j_{eff}^{\nu},
\end{equation}
where
\begin{equation}
\begin{split}
\mathcal{S}^{\mu\nu} & =F^{\delta\nu}\frac{1}{\Box}(F^{\mu\beta}F_{\delta\beta})-F^{\delta\mu}\frac{1}{\Box}(F^{\nu\beta}F_{\delta\beta})+F^{\kappa\nu}\frac{1}{\Box}(F_{\kappa}^{ \ \beta}F_{ \ \beta}^{\mu})-F^{\kappa\mu}\frac{1}{\Box}(F_{\kappa}^{ \ \beta}F_{ \ \beta}^{\nu})\\
 & +F_{ \ \sigma}^{\mu}F^{\delta\sigma}\frac{1}{\Box}F_{\delta}^{ \ \nu}-F_{ \ \sigma}^{\nu}F^{\delta\sigma}\frac{1}{\Box}F_{\delta}^{ \ \mu}+F_{ \ \sigma}^{\kappa}F^{\mu\sigma}\frac{1}{\Box}F_{\kappa}^{ \ \nu}-F_{ \ \sigma}^{\kappa}F^{\nu\sigma}\frac{1}{\Box}F_{\kappa}^{ \ \mu}\\
 & -F^{\mu\nu}\frac{1}{2\Box}(F^{\alpha\beta}F_{\alpha\beta})-F_{\alpha\beta}F^{\alpha\beta}\frac{1}{2\Box}F^{\mu\nu},
\end{split}
\end{equation}
and the quantity $j_{eff}^{\nu}=\frac{\kappa^{2}}{2}(\partial_{\mu}\mathcal{S}^{\mu\nu})$ can effectively be treated as a source current. The equation (\ref{source}) is similar to Maxwell's equation in the presence of a conserved current which originates from the coupling of gauge field with matter fields. However, in the equation (\ref{source}), the conserved current $j_{eff}^{\nu}$ depends only on the gauge field. This current vanishes in the absence of gravity ($\kappa\rightarrow0$). Using the definitions of electric and magnetic fields $F^{0i}=E^{i}$ and $F^{ij}=\epsilon^{ijk}B^{k}$, it can be shown that
\begin{equation}\label{eqn.3}
\begin{split}
\mathcal{S}^{0i} & =2B^{k}\frac{1}{\Box}(E^{i}B^{k})-2B^{k}\frac{1}{\Box}(E^{k}B^{i})-2E^{j}\frac{1}{\Box}(E^{i}E^{j})-2E^{j}\frac{1}{\Box}(B^{i}B^{j})\\
 & +2E^{i}\left(\vec{B}.\frac{1}{\Box}\vec{B}\right)-2B^{i}\left(\vec{E}.\frac{1}{\Box}\vec{B}\right)-2E^{i}\left(\vec{E}.\frac{1}{\Box}\vec{E}\right)-3B^{i}\left(\vec{B}.\frac{1}{\Box}\vec{E}\right)+(\vec{E}^{2}-\vec{B}^{2})\frac{1}{\Box}E^{i}.
\end{split}
\end{equation}
This brings out the first set of modified Maxwell's equations in the presence of gravitons, which follows from (\ref{eqn.3})
\begin{equation}
\vec{\nabla}.\vec{D}=0, \ \vec{D}=\vec{E}-\vec{\mathcal{S}},\ D^{i}=F^{0i}-\mathcal{S}^{0i}.
\end{equation}
Similarly for the spatial indices $(ij)$, we define the quantity
\begin{equation}
\begin{split}
\tilde{\mathcal{S}}^{m} & =\epsilon^{ijm}\mathcal{S}^{ij}\\
\implies \vec{\tilde{\mathcal{S}}} & =4\vec{B}\frac{1}{\Box}(\vec{E}^{2}+\vec{B}^{2})+4\vec{E}^{2}\frac{1}{\Box}\vec{B}+4\left(\vec{E}\times\frac{1}{\Box}(\vec{E}\times\vec{B})\right)-2\vec{E}\times\left((\frac{1}{\Box}\vec{B})\times\vec{E}\right)-(\vec{E}\times\vec{B})\times\frac{1}{\Box}\vec{E}.
\end{split}
\end{equation} 
Therefore,
\begin{equation}
\vec{\nabla}\times\vec{H}=\frac{\partial\vec{D}}{\partial t}, \ \vec{H}=\vec{B}-\vec{\tilde{\mathcal{S}}}.
\end{equation}

The above set of equations bring out that in the presence of the gravitons, the photon medium gets polarized with $\vec{\mathcal{S}}$ and gains magnetization, denoted by $\vec{\tilde{\mathcal{S}}}$. These are non-linear features that could be useful for detecting GW. On the other hand, in the case of massive gravitons, each $\frac{1}{\Box}$ term gets modified to $\frac{1}{\Box-m^{2}}$. This set of non-linear Maxwell equations can put further constraints on the mass of graviton by a comparison of the experimental data with the theoretical prediction.

In a birefringent medium, uniform plane waves can be decomposed into two orthogonal polarization states that propagate at two different speeds. These two states develop a phase difference as they propagate, which alters the total polarization of the wave. In this case, both the vectors $\vec{\mathcal{S}}, \ \vec{\tilde{\mathcal{S}}}$ depend on electric and magnetic fields $(\vec{E},\vec{B})$ nonlinearly. As a result, the permittivities, permeabilities, and refractive indices of this anisotropic medium strongly depend on the $\vec{E},\vec{B}$ fields non-linearly. This characterizes the birefringence property of the vacuum \cite{hattori2013vacuum, ataman2018vacuum, nakamiya2017probing, Kruglov_2015}. In the case of non-vanishing sources, these quantities also carry information about the physical properties of the compact objects, as discussed above (this follows from the linearity property of the on-shell equation (\ref{longitudinal pol}) and (\ref{eqn.2})). Therefore, the above set of equations can be used not only in the detection process but also to extract information about compact objects like neutron stars, white-dwarfs, binary mergers. Since in the case of massive gravity theory, the mass of the graviton is involved in the $j_{eff}^{\mu}$ current, hence, the permittivities, refractive indices, and permeabilities depend on the mass of gravitons. As a result of this, measurements on these quantities put a bound on the mass of gravitons.

\section{Scattering process between photons in the presence of gravitons}
\subsection{Action in momentum space}
Since the detection of a single graviton is very challenging from the perspective of the present technology, our approach of integrating out the graviton degrees of freedom and writing an effective action for photons that takes into account the interactions between photons and gravitons would be helpful since there have been impressive advances in the field of photon detection. Now our aim is to write down the interacting part of the action in momentum space from which the scattering amplitudes can be calculated. Here also we do not assume any external source producing GW.
The interaction part of the action in momentum space takes the following form
\begin{equation}
(2\pi)^{4}\int\prod_{i=1}^{4}d^{4}k_{i}\delta^{(4)}(k_{1}+k_{2}+k_{3}+k_{4})A^{\mu}(k_{1})A^{\nu}(k_{2})A^{\rho}(k_{3})A^{\sigma}(k_{4})\mathcal{V}_{\mu\nu\rho\sigma}(k_{1},k_{2},k_{3},k_{4}),
\end{equation}
where
\begin{equation}
\begin{split}
\mathcal{V}_{\mu\nu\rho\sigma}(k_{1},k_{2},k_{3},k_{4}) & =\Big[k_{1}.k_{2}k_{3}.k_{4}\eta_{\mu\nu}\eta_{\rho\sigma}-k_{1}.k_{2}\eta_{\mu\nu}k_{3\sigma}k_{4\rho}-k_{3}.k_{4}\eta_{\rho\sigma}k_{1\nu}k_{2\mu}+k_{1\nu}k_{2\mu}k_{3\sigma}k_{4\rho}-k_{1}.k_{2}k_{3}.k_{4}\eta_{\mu\rho}\eta_{\nu\sigma}\\
+k_{1}.k_{2}\eta_{\mu\rho}k_{3\sigma}k_{4\nu} & +k_{1}.k_{2}\eta_{\nu\sigma}k_{3\mu}k_{4\rho}-k_{1}.k_{2}\eta_{\rho\sigma}k_{3\mu}k_{4\nu}+k_{3}.k_{4}\eta_{\mu\rho}k_{1\nu}k_{2\sigma}-k_{2}.k_{4}\eta_{\mu\rho}k_{1\nu}k_{3\sigma}-k_{1\nu}k_{2\sigma}k_{3\mu}k_{4\rho}\\
+k_{2}.k_{4}\eta_{\rho\sigma}k_{1\nu}k_{3\mu} & +k_{3}.k_{4}\eta_{\nu\sigma}k_{1\rho}k_{2\mu}-k_{1\rho}k_{2\mu}k_{3\sigma}k_{4\nu}-k_{1}.k_{3}\eta_{\nu\sigma}k_{2\mu}k_{4\rho}+k_{1}.k_{3}\eta_{\rho\sigma}k_{2\mu}k_{4\nu}-k_{3}.k_{4}\eta_{\mu\nu}k_{1\rho}k_{2\sigma}\\
+k_{2}.k_{4}\eta_{\mu\nu}k_{1\rho}k_{3\sigma} & +k_{1}.k_{3}\eta_{\mu\nu}k_{2\sigma}k_{4\rho}-\eta_{\mu\nu}\eta_{\rho\sigma}k_{1}.k_{3}k_{2}.k_{4}\Big](2\pi)^{4}\frac{\kappa^{2}}{(k_{3}+k_{4})^{2}}\delta^{(4)}(k_{1}+k_{2}+k_{3}+k_{4}).
\end{split}
\end{equation}
Here we compute the scattering amplitudes using perturbative technique around the free-field theory, namely the free Maxwell theory in which photons are massless. Hence, the scattering amplitudes are computed considering two polarization states of photons. Let us denote the polarization tensors of the photons as $\varepsilon_{i}\equiv\varepsilon(k_{i},\lambda_{i})$, where $\lambda_{i}\in\{1,2\}$ which satisfy
\begin{equation}
\sum_{\lambda=1,2}\varepsilon_{i}^{(\lambda)}\varepsilon_{j}^{(\lambda)}=\left(\delta_{ij}-\frac{k^{i}k^{j}}{|\vec{k}|^{2}}\right).
\end{equation}
However, if we take into account the quantum corrections by considering one-loop self-energy diagrams (shown in section \ref{quantum correction}), then we need to consider the massive spin-1 degrees of freedom in the scattering amplitude computation.

\subsection{Vacuum to 4-photons scattering amplitude}
In this section, we compute the scattering amplitude of a process in which from the vacuum, four photons are produced with the same polarization $\lambda=1$. The scattering amplitude is defined by $\bra{f}S^{(1)}\ket{i}$ with $S^{(1)}$ interaction term in the action and $\ket{i}=\ket{0}$ is the initial state and the final state is $\prod_{i=1}^{4}\hat{a}^{\dagger}(k_{i},\lambda_{i})\ket{0}$ and the polarizations $\lambda_{1},\lambda_{2},\lambda_{3},\lambda_{4}$ can take any value. The corresponding scattering amplitude would be
\begin{equation}
\begin{split}
\mathcal{M}_{2} & =\sum_{\tilde{\lambda}_{1},\tilde{\lambda}_{2},\tilde{\lambda}_{3},\tilde{\lambda}_{4}}\int \prod_{i=1}^{4}d^{4}p_{i}\mathcal{V}_{\mu\nu\rho\sigma}(p_{1},p_{2},p_{3},p_{4})\bra{0}\hat{a}(k_{4},\lambda_{4})\hat{a}(k_{3},\lambda_{3})\hat{a}(k_{2},\lambda_{2})\hat{a}(k_{1},\lambda_{1})\\
\times & \hat{a}^{\dagger}(p_{1},\tilde{\lambda}_{1})\hat{a}^{\dagger}(p_{2},\tilde{\lambda}_{2})\hat{a}^{\dagger}(p_{3},\tilde{\lambda}_{2})\hat{a}^{\dagger}(p_{4},\tilde{\lambda}_{4})\ket{0}\varepsilon^{\mu}(p_{1},\tilde{\lambda}_{1})\varepsilon^{\nu}(p_{2},\tilde{\lambda}_{2})\varepsilon^{\rho}(p_{3},\tilde{\lambda}_{3})\varepsilon^{\sigma}(p_{4},\tilde{\lambda}_{4})\\
 & =\sum_{\tilde{\lambda}_{1},\tilde{\lambda}_{2},\tilde{\lambda}_{3},\tilde{\lambda}_{4}}\sum_{p\in S_{4}}\mathcal{V}_{\mu\nu\rho\sigma}(k_{p(1)},k_{p(2)},k_{p(3)},k_{p(4)})\varepsilon^{\mu}(k_{p(1)},\lambda_{p(1)})\varepsilon^{\nu}(k_{p(2)},\lambda_{p(2)})\varepsilon^{\rho}(k_{p(3)},\lambda_{p(3)})\\ 
 & \times\varepsilon^{\sigma}(k_{p(4)},\lambda_{p(4)})\delta_{\tilde{\lambda}_{1},\lambda_{p(1)}}\delta_{\tilde{\lambda}_{2},\lambda_{p(2)}}\delta_{\tilde{\lambda}_{3},\lambda_{p(3)}}\delta_{\tilde{\lambda}_{4},\lambda_{p(4)}}.
\end{split}
\end{equation} 
\subsection{Scattering amplitude of decay process}
Now we consider $1\rightarrow3$ particle decay process where initial state is $\ket{i}=\hat{a}^{\dagger}(k_{1},\lambda_{1})\ket{0}$ and final state is $\ket{f}=\hat{a}^{\dagger}(k_{2},\lambda_{2})\hat{a}^{\dagger}(k_{3},\lambda_{3})\hat{a}^{\dagger}(k_{4},\lambda_{4})\ket{0}$, for which we can write $\bra{f}S^{(1)}\ket{i}=\mathcal{M}$
where
\begin{equation}
\begin{split}
\mathcal{M} & =\sum_{\tilde{\lambda}_{1},\tilde{\lambda}_{2},\tilde{\lambda}_{3},\tilde{\lambda}_{4}}\int \prod_{i=1}^{4}d^{4}p_{i}\mathcal{V}_{\mu\nu\rho\sigma}(p_{1},p_{2},p_{3},p_{4})\varepsilon^{\mu}(p_{1},\tilde{\lambda}_{1})\varepsilon^{\nu}(p_{2},\tilde{\lambda}_{2})\varepsilon^{\rho}(p_{3},\tilde{\lambda}_{3})\varepsilon^{\sigma}(p_{4},\tilde{\lambda}_{4})\\
 & \times\Bigg[4\Big[\sum_{P\in S_{3}}\delta^{(4)}(p_{1}-k_{1})\delta^{(4)}(p_{2}-k_{P(2)})\delta^{(4)}(p_{3}-k_{P(3)})\delta^{(4)}(p_{4}-k_{P(4)})\delta_{\tilde{\lambda}_{1},\lambda_{1}}\delta_{\tilde{\lambda}_{2},\lambda_{p(2)}}\delta_{\tilde{\lambda}_{3},\lambda_{p(3)}}\delta_{\tilde{\lambda}_{4},\lambda_{p(4)}}\\
 & +\sum_{m,n=2}^{4}\delta^{(4)}(p_{1}-k_{m})\delta^{(4)}(k_{1}-p_{n})\delta_{\tilde{\lambda}_{1},\lambda_{m}}\delta_{\tilde{\lambda}_{n},\lambda_{1}}\prod_{m'\neq m,n'\neq n\in\{1,2,3\}}\delta^{(4)}(k_{m'}-p_{n'})\delta_{\tilde{\lambda}_{n'},\lambda_{m'}}\Big]\\
 & -3\delta_{\tilde{\lambda}_{2},\tilde{\lambda}_{1}}\delta^{(4)}(p_{1}-p_{2})\Big[\sum_{m=2,3,4}\delta_{\lambda_{1},\lambda_{m}}\delta^{(4)}(k_{1}-k_{m})(\delta_{\tilde{\lambda}_{4},\lambda_{m_{>}}}\delta^{(4)}(p_{4}-m_{>})\delta_{\tilde{\lambda}_{3},\lambda_{m_{<}}}\delta^{(4)}(p_{3}-m_{<})\\
 & +\delta_{\tilde{\lambda}_{3},\lambda_{m_{>}}}\delta^{(4)}(p_{3}-m_{>})\delta_{\tilde{\lambda}_{4},\lambda_{m_{<}}}\delta^{(4)}(p_{4}-m_{<}))\Big]\\
 & -2\delta_{\tilde{\lambda}_{3},\tilde{\lambda}_{1}}\delta^{(4)}(p_{1}-p_{3})\Big[\sum_{m=2,3,4}\delta_{\lambda_{1},\lambda_{m}}\delta^{(4)}(k_{1}-k_{m})(\delta_{\tilde{\lambda}_{4},\lambda_{m_{>}}}\delta^{(4)}(p_{4}-m_{>})\delta_{\tilde{\lambda}_{2},\lambda_{m_{<}}}\delta^{(4)}(p_{2}-m_{<})\\
\end{split}
\end{equation}
\begin{align*}
 & +\delta_{\tilde{\lambda}_{2},\lambda_{m_{>}}}\delta^{(4)}(p_{2}-m_{>})\delta_{\tilde{\lambda}_{4},\lambda_{m_{<}}}\delta^{(4)}(p_{4}-m_{<}))\Big]\\
 & -\delta_{\tilde{\lambda}_{4},\tilde{\lambda}_{1}}\delta^{(4)}(p_{1}-p_{4})\Big[\sum_{m=2,3,4}\delta_{\lambda_{1},\lambda_{m}}\delta^{(4)}(k_{1}-k_{m})(\delta_{\tilde{\lambda}_{4},\lambda_{m_{>}}}\delta^{(4)}(p_{4}-m_{>})\delta_{\tilde{\lambda}_{1},\lambda_{m_{<}}}\delta^{(4)}(p_{1}-m_{<})\\
 & +\delta_{\tilde{\lambda}_{1},\lambda_{m_{>}}}\delta^{(4)}(p_{1}-m_{>})\delta_{\tilde{\lambda}_{4},\lambda_{m_{<}}}\delta^{(4)}(p_{4}-m_{<}))\Big]\Bigg]\times(2\pi)^{4}\delta^{(4)}(p_{1}-p_{2}-p_{3}-p_{4})\prod_{i=1}^{4}\frac{1}{\sqrt{2\omega_{i}}(2\pi)^{\frac{3}{2}}}
\end{align*}
where if $m=2$ then $m_{<}=3,m_{>}=4$, if $m=3$ then $m_{<}=2,m_{>}=4$ and if $m=4$ then $m_{<}=2,m_{>}=3$ and $m',n'$ can take one value at one time only.
\subsection{Features of scattering amplitudes}
Up to one-loop, all scattering amplitudes shown above are proportional to $\kappa^{2}=\frac{16\pi G}{c^{4}}$ which is $\mathcal{O}(10^{-43})$. This is very small in magnitude for measurement in any scattering experiment. However, using weak value amplification \cite{nishizawa2015weak, li2017adaptive, carrasco2019weak} this magnitude can be amplified through suitable pre-selected and post-selected scattering states, shown below.

Another interesting aspect of these scattering processes is that the helicity of photons through these processes is not conserved. This was first shown in \cite{weber1962interaction}, and the reason behind this will be discussed below.

The other feature found is that the vertex function $\mathcal{V}_{\mu\nu\rho\sigma}(k_{1},k_{2},k_{3},k_{4})$ contains a factor of $\frac{1}{(k_{3}+k_{4})^{2}}$ which shows an IR pole at $k_{3}+k_{4}=0$ that can be avoided by adding soft photons \cite{frye2019infrared, kapec2017infrared}. This also happens in QED, although their vertex function does not have any poles in momentum space. Note that this feature is in-built in photons due to interaction with massless gravitons since asymptotically the interaction term in the action is non-zero for soft gravitons, but this is not the case for massive gravitons. Therefore, one would expect the absence of the IR pole for massive gravitons. This is indeed the case as there we need to replace $\frac{1}{(k_{3}+k_{4})^{2}}$ by $\frac{1}{(k_{3}+k_{4})^{2}+m^{2}}$.

\subsection{Duality symmetry}
One of the beautiful features of Maxwell equations in free-space is that it has the symmetry of exchanging electric and magnetic fields. Maxwell equations in free-space are symmetric under the following continuous transformation, known as dual transformation \cite{li2001alternative, figueroa1998electromagnetic, aschieri2008three, galvao2000consistent, donev2000new}
\begin{equation}
\vec{E}\rightarrow\vec{E}\cos\theta+\vec{B}\sin\theta, \ \vec{B}\rightarrow\vec{B}\cos\theta-\vec{E}\sin\theta,
\end{equation}
where $\theta$ is an arbitrary angle.

A concrete analysis of conserved quantity (Noether's charge) corresponding to this continuous symmetry revealed that the pertinent pseudo scalar integrated over a spatial hypersurface represents the difference between the number of left- and right-hand circularly polarized photons which is nothing but optical helicity. Hence, this duality symmetry leads to the conservation of helicity of light \cite{crimin2019optical, fernandez2015helicity, trueba1996electromagnetic, afanasiev1996helicity, candlin1965analysis}.

In standard Maxwell's electromagnetism
\begin{equation}
\mathcal{L}=-\frac{1}{4}F_{\alpha\beta}F^{\alpha\beta}=\frac{1}{2}(\vec{E}^{2}-\vec{B}^{2}),
\end{equation}
corresponds to free Maxwell equations
\begin{equation}
\begin{split}
\partial_{\alpha}F^{\alpha\beta} & =0, \ \partial_{\alpha}*F^{\alpha\beta}=0, \ *F^{\alpha\beta}=\frac{1}{2}\varepsilon^{\alpha\beta\gamma\delta}F_{\gamma\delta}.
\end{split}
\end{equation}
It is important to note that coupling of matter with photon breaks this symmetry. If Lagrangian acquires a term $j^{\alpha}A_{\alpha}$ which implies equations of motion become
\begin{equation}
\partial_{\alpha}F^{\alpha\beta}=-j_{E}^{\beta}, \ \partial_{\alpha}*F^{\alpha\beta}=-j_{M}^{\beta} \ (\text{magnetic current}).
\end{equation}
Now we define two Lorentz invariant quantities
\begin{equation}
\begin{split}
I_{1} & =-\frac{1}{2}F_{\alpha\beta}F^{\alpha\beta}=\vec{E}^{2}-\vec{B}^{2}, \ I_{2}=-\frac{1}{2}*F^{\alpha\beta}F_{\alpha\beta}=2\vec{E}.\vec{B},
\end{split}
\end{equation}
which are important for the subsequent discussion. Note that
\begin{equation}
\mathcal{L}_{\text{free}}=\frac{1}{2}\mathfrak{R}(\vec{R}.\vec{R})=\frac{1}{2}I_{1}.
\end{equation}
Under the dual transformation
\begin{equation}
\begin{split}
I_{1}\rightarrow & I_{1}\cos2\theta+I_{2}\sin2\theta, \ I_{2}\rightarrow I_{2}\cos2\theta-I_{1}\sin2\theta,
\end{split}
\end{equation}
the Lagrangian density transforms as
\begin{equation}\label{eqn.4}
\mathcal{L}_{\text{free}}\rightarrow\mathcal{L}_{\text{free}}\cos2\theta-\frac{1}{4}*F^{\alpha\beta}F_{\alpha\beta}\sin2\theta.
\end{equation}
Although the above transformation changes Lagrangian density, the Maxwell equations remain unchanged since
\begin{equation}
*F^{\alpha\beta}F_{\alpha\beta}=2\partial_{\alpha}(*F^{\alpha\beta}A_{\beta}),
\end{equation}
where we have used the on-shell relation $\partial_{\alpha}*F^{\alpha\beta}=0$ for free-space. Considering the infinitesimal version of the transformation (\ref{eqn.4})
\begin{equation}
\mathcal{L}_{\text{free}}\rightarrow\mathcal{L}_{\text{free}}-\theta\partial_{\alpha}(*F^{\alpha\beta}A_{\beta}),
\end{equation}
it can be seen that $\mathcal{L}_{\text{free}}$ changes only by a total derivative which therefore is a symmetry transformation and 
\begin{equation}
J^{\alpha}=*F^{\alpha\beta}A_{\beta},
\end{equation}
is the corresponding conserved charge in the absence of sources. In the presence of sources,
\begin{equation}
\begin{split}
\partial_{\alpha}J^{\alpha} & =\partial_{\alpha}*F^{\alpha\beta}A_{\beta}+\frac{1}{2}*F^{\alpha\beta}F_{\alpha\beta}\\
 & =-j_{M}^{\alpha}A_{\alpha}-I_{2}\neq0.
\end{split}
\end{equation}
$J^{\alpha}$ is no longer a conserved current. In the presence of gravitons we have a non-zero source which is $j_{eff}^{\nu}=\frac{\kappa^{2}}{2}\partial_{\mu}\mathcal{S}^{\mu\nu}$, (see (\ref{source})), which guarantees that $\vec{E}.\vec{B}\neq0$, thereby bringing out the violation of helicity conservation \cite{cameron2012optical, fernandez2013necessary}.
\section{Weak measurements}
Each physical quantity $A$ in quantum mechanics can be described by a Hermitian operator $\hat{A}$ in the Hilbert space of a quantum system $S$. Ideal (or projective or strong) measurements of that system S are known as the projection postulate. An arbitrary state of the system is in general not an eigenstate of the observable $\hat{A}$, but rather a linear superposition of a complete orthonormal basis states $\ket{a_{n}}$ with $\ket{\psi}=\sum_{n}\psi_{n}\ket{a_{n}}$. The interaction Hamiltonian is $\hat{H}=\chi\hat{p}\otimes\hat{A}$ with the interaction strength $|\chi|\ll1$, where $\hat{p}$ is the momentum operator of the device and $\hat{A}$ is the operator whose expectation value needs to be measured. An ideal measuring device possess well-defined initial ($p_i$) and final ($p_f$) values of momentum with $\Delta p$ as the width (described by a gaussian state $\ket{\phi_{i}}$). The difference ($p_f-p_i$), is the device's pointer reading, which indicates the value of $A$. If $\Delta p$ is much larger than the spread of the eigenvalues $\{a_{n}\}$ then it is in the ``weak measurement" domain \cite{svensson2013pedagogical, tamir2013introduction, duck1989sense, ritchie1991realization, PhysRevLett.60.1351}. Aharonov, Bergmann, and Lebowitz defined pre- and post-selected measurements. An ensemble of quantum systems is pre-selected in the state $\ket{\psi_{i}}$. All ensemble members have gone through a measurement of the observable $A$ and it may be a weak or strong measurement. Finally, a projective strong measurement is performed on the ensemble. The final measurement is projective of a variable with a discrete, non-degenerate spectrum. That final measurement leaves the system in one of the orthogonal states. Hence, the possible outcome of the measurement $A$ is a function of both the pre- and post-selected states of the system. This measurement procedure is known as a pre- and post-selected (PPS) measurement.

After post selection with the pre-selected state $\ket{\psi_i}=\sum_{n}\psi_{n}\ket{a_{n}}$, the state of measuring device will be:
\begin{equation}
\ket{\phi_f}\approx \langle\psi_f\ket{\psi_i}  \int e^{-\Delta^2(p-\chi\langle\hat{A}\rangle_{w})^2}\ket{p}dp.
\end{equation} 
The device measures the quantity  
$\langle\hat{A}\rangle_{w}\equiv\frac{\bra{\psi_f}\hat{A}\ket{\psi_{i}}}{\braket{\psi_f|\psi_{i}}}$,
known as the weak value \cite{ogawa2019framework, dass2017optimal} where $\langle\hat{A}\rangle_{w}\ll\frac{1}{\Delta}$ and $\ket{\psi_{f}}$ is the post-selected state. This is the Aharanov-Albert-Vaidman (AAV) limit.

\subsection{Measuring scattering amplitudes using weak measurement protocol}
In the scattering amplitudes we are interested in measuring the quantity $\braket{f|i}$ where $\ket{i}$ is the initial state at past infinity $t\rightarrow-\infty$ and $\ket{f}$ is the final state at future infinity at $t\rightarrow\infty$ or in the    Schr$\ddot{o}$dinger picture $\bra{f}e^{-i\hat{H}T}\ket{i}|_{T\rightarrow\infty}=\bra{f}\hat{\mathcal{S}}\ket{i}$. Therefore, if we choose the operator $\hat{A}$ in weak measurement protocol to be the scattering matrix $\hat{S}$, and choose the initial state to be a many-particle state $\ket{i}=\ket{k_{1},k_{2},\ldots,k_{m}}$ and a post-selected state  $\ket{f}=(1-\epsilon)\ket{l_{1},\ldots,l_{n}}+\sqrt{2\epsilon}\ket{k_{1},k_{2},\ldots,k_{m}}$ such that $\epsilon\ll1$, we have
\begin{equation}
\begin{split}
\langle\hat{A}\rangle_{w}=\langle\hat{\mathcal{S}}\rangle_{w} & =\frac{(1-\epsilon)\bra{l_{1},\ldots,l_{n}}\hat{\mathcal{S}}\ket{k_{1},\ldots,k_{m}}+\sqrt{2\epsilon}\bra{k_{1},\ldots,k_{m}}\hat{\mathcal{S}}\ket{k_{1},\ldots,k_{m}}}{\sqrt{2\epsilon}}\\
 & \approx\frac{1}{\sqrt{2\epsilon}}\bra{l_{1},\ldots,l_{n}}\hat{\mathcal{S}}\ket{k_{1},\ldots,k_{m}},
\end{split}
\end{equation}
which is a scattering amplitude of $m$-particle state with momenta $k_{1},\ldots,k_{m}$ to $n$-particle state with momenta $l_{1},\ldots,l_{n}$ and an amplifying factor $\frac{1}{\sqrt{2\epsilon}}$. This facilitates the measurement of scattering amplitude or cross-sections of any process in the theory through the weak measurement protocol by suitably choosing a channel with a collection of particular initial and post-selected states. Using the cascaded weak measurement strategy \cite{hu2017cascaded}, $\langle\hat{\mathcal{S}}\rangle_{w}$ can be amplified by $\mathcal{O}(10^{12})$. The above weak-amplification holds as long as the pre- and post-selected states are not completely orthogonal to each other, as mentioned above. Thus, separating out such pre- and post-selected states play a vital role in weak-measurement. Because of this amplification of scattering amplitudes due to the weak measurement, probing the features of scattering processes becomes relatively easier. Hence, the weak measurement of scattering amplitudes in the scattering processes between photons would be useful in the detection of GW interacting with light \cite{hu2017gravitational}. The crucial role of GW  in probing Cosmology was pointed out in \cite{caprini2018cosmological,davies2014quantum}. This puts into perspective our proposal that weak measurements on astrophysical and cosmological photons interacting with GW improve the detection of GW significantly.

\section{Effective action}\label{quantum correction}
In order to take into account the quantum corrections, quantum effective action is obtained here by a one-loop computation.
\subsection{Self-energy of photons}
\begin{figure}
\begin{center}
\includegraphics[height=3cm, width=5cm]{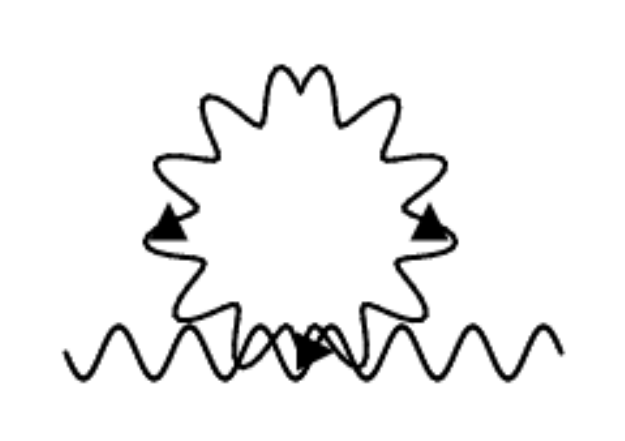}
\end{center}
\caption{1-loop self-energy diagram}
\end{figure}
Figure.1 depicts the one-loop self-energy diagram, computed using Feynman diagrammatic techniques and equal to
\begin{equation}
-i\int\frac{d^{4}k}{(2\pi)^{4}}\mathcal{V}_{\mu\nu\rho\sigma}(p,k,k,p)\frac{\eta^{\nu\rho}}{k^{2}+\mu^{2}},
\end{equation}
where $\mu$ is mass-regulator of photons. This implies that the diagram can be mathematically represented by the following expression
\begin{equation}\label{eqn.6}
\begin{split}
-i\kappa^{2}\int\frac{d^{4}k}{(2\pi)^{4}} & \frac{1}{k^{2}+\mu^{2}}\frac{1}{(p+k)^{2}}\Big[-(p.k)^{2}\eta_{\mu\sigma}-p^{2}k_{\mu}k_{\sigma}+p.kk_{\mu}p_{\sigma}+p.kp_{\mu}k_{\sigma}\Big]\\
=-\kappa^{2}[(p^{2})^{2}\eta_{\mu\sigma} & -p^{2}p_{\mu}p_{\sigma}]\frac{1}{(4\pi)^{2}}\Bigg[\left(\frac{2}{\varepsilon}+\psi(0)\right)\int_{0}^{1}dx \ x^{2}\left(1-\frac{\varepsilon}{2}\ln\left(\frac{p^{2}x(1-x)+\mu^{2}(1-x)}{4\pi\Lambda^{2}}\right)\right)\\
 & +\left(\frac{2}{\varepsilon}-\psi(1)\right)\int_{0}^{1}dx \ x(1-x)\left(1-\frac{\varepsilon}{2}\ln\left(\frac{p^{2}x(1-x)+\mu^{2}(1-x)}{4\pi\Lambda^{2}}\right)\right)\Bigg],
\end{split}
\end{equation}
where $\Gamma, \ \psi$ denote Gamma and Euler's function, respectively. $\Lambda$ is the momentum scale (effective scale) up to which this theory is valid and $\varepsilon=4-D$.
  
We can now safely take $\mu=0$ since there is no IR divergence when $p=0$. From the above expression, the divergent part can be omitted by adding suitable counterterms and we are left with a finite part whose contribution is
\begin{equation}\label{eqn.7}
\begin{split}
-\kappa^{2}[(p^{2})^{2}\eta_{\mu\sigma} & -p^{2}p_{\mu}p_{\sigma}]\frac{1}{(4\pi)^{2}}\left(\mathcal{C}-\frac{1}{2}\ln\frac{p^{2}}{4\pi\Lambda^{2}}\right), \ \mathcal{C}=\frac{\psi(0)}{3}-\frac{\psi(1)}{6}+1.
\end{split}
\end{equation}
Therefore, the quadratic part of the effective Lagrangian density of photon degrees of freedom (after integrating out the graviton degrees of freedom) up to one-loop becomes
\begin{equation}
\begin{split}
\mathcal{L}_{eff}^{(2)} & =-\frac{1}{4}F_{\mu\nu}F^{\mu\nu}+\kappa^{2}A^{\mu}\Big[\left(\alpha-\alpha'\ln\frac{-\Box}{4\pi\Lambda^{2}}\right)\Box\partial_{\mu}\partial_{\nu}-\left(\alpha-\alpha'\ln\frac{-\Box}{4\pi\Lambda^{2}}\right)\eta_{\mu\nu}\Box^{2}\Big]A^{\nu}\\
 & \equiv\frac{1}{2}A_{\mu}\Bigg[(\Box\eta^{\mu\nu}-\partial^{\mu}\partial^{\nu})+\kappa^{2}\Big[\left(\alpha-\alpha'\ln\frac{-\Box}{4\pi\Lambda^{2}}\right)\Box\partial^{\mu}\partial^{\nu}-\left(\alpha-\alpha'\ln\frac{-\Box}{4\pi\Lambda^{2}}\right)\eta^{\mu\nu}\Box^{2}\Big]\Bigg]A_{\nu}.
\end{split}
\end{equation}
Note that taking into account the quantum correction of photons up to one-loop generates the non-local $\ln\frac{-\Box}{4\pi\Lambda^{2}}$ term in the quadratic part of the effective action with $\alpha=\frac{\mathcal{C}}{(4\pi)^{2}}$ and $\alpha'=\frac{1}{2(4\pi)^{2}}$ (follows from transversality condition). A similar kind of non-local term is recently found in effective field theory GR in \cite{calmet2016gravitational, calmet2018gravitational}.

We now calculate the dispersion relation (on-shell) due to quantum corrections which take into account effective interactions with gravitons. But before that, we need to choose a gauge and in this case, we choose the Lorentz gauge $\partial_{\mu}A^{\mu}=0$, due to which the effective Lagrangian density up to quadratic part becomes
\begin{equation}
\mathcal{L}_{eff}^{(2)}=\frac{1}{2}A_{\mu}\Bigg[\Box\eta^{\mu\nu}-\kappa^{2}\left(\alpha-\alpha'\ln\frac{-\Box}{4\pi\Lambda^{2}}\right)\eta^{\mu\nu}\Box^{2}\Bigg]A_{\nu}.
\end{equation} 
Therefore, the dispersion relation of photons becomes
\begin{equation}
k^{2}\left(1+\kappa^{2}\left(\alpha-\alpha'\ln\frac{k^{2}}{4\pi\Lambda^{2}}\right)k^{2}\right)=0,
\end{equation}
which has 2 branches. One of them is usual photon dispersion relation in free-field theory $k^{2}=0$ and other is the non-trivial scale dependent dispersion relation 
\begin{equation}
1+\kappa^{2}\left(\alpha-\alpha'\ln\frac{k^{2}}{4\pi\Lambda^{2}}\right)k^{2}=0.
\end{equation}
Thus, the above dispersion relation becomes
\begin{equation}
k^{2}=4\pi\Lambda^{2}e^{2\mathcal{C}+\mathcal{W}_{\pm1}\left(\frac{12\pi}{\kappa^{2}\Lambda^{2}}\right)},
\end{equation}
where $\mathcal{W}_{\pm1}(x)$ is the W-Lambert function, which has a non-zero imaginary part that shows that photon amplitude decays exponentially in time. A similar dispersion relation for gravitons in the effective field theory of GR was recently found in \cite{calmet2016gravitational}, \cite{calmet2018gravitational}. The existence of massive photons is shown here without the non-minimal coupling of photons to curvature or dark-energy \cite{Kouwn_2016}. Scale dependence of this dispersion in low-energy theory implies large-scale anomalies that are consistent with the observations \cite{Adhikari_2016, rassat2014planck}. Further, the non-zero imaginary part of $\mathcal{W}_{\pm1}(x)$ implies that photons are unstable in nature and decay into other photons which explains why the CMB spectrum would not fit the near-perfect thermal curve \cite{diacoumis2017using}. Further, the real part of $\mathcal{W}_{\pm1}(x)$ also implies that photons interacting with gravitons, also have longitudinal polarization.

However, the presence of graviton mass gives one additional scale which is completely independent of the Planck length scale $\kappa$ and this new scale depends on the Cosmological constant \cite{haranas2014mass, liu2004gravitational}. In principle, this could also come self-energy contribution to the graviton field from the inflaton field \cite{lin2013massive, lin2016resonant} which leads to inflation of the Universe at a very early stage. In order to see how this new scale emerges into effective action at the quadratic level, we need to replace in (\ref{eqn.6}) $(p+k)^{2}\rightarrow(p+k)^{2}+m^{2}$ which modifies (\ref{eqn.7}), the finite part
\begin{equation}
\begin{split}
=-\kappa^{2}[(p^{2})^{2}\eta_{\mu\sigma} & -p^{2}p_{\mu}p_{\sigma}]\frac{1}{(4\pi)^{2}}\Bigg[\tilde{\mathcal{C}}-\left(\frac{1}{2}+\frac{3m^{2}}{2p^{2}}+\frac{3m^{4}}{2p^{4}}+\frac{m^{6}}{2p^{6}}\right)\ln\left(\frac{p^{2}+m^{2}}{4\pi\Lambda^{2}}\right)\\
+\frac{3m^{2}}{2p^{2}} & \left(\ln\left(\frac{m^{2}}{4\pi\Lambda^{2}}\right)+1\right)+\frac{3m^{4}}{2p^{4}}\left(\ln\left(\frac{m^{2}}{4\pi\Lambda^{2}}\right)+\frac{1}{3}\right)+\frac{1}{2}\frac{m^{6}}{p^{6}}\ln\left(\frac{m^{2}}{4\pi\Lambda^{2}}\right)\Bigg],
\end{split}
\end{equation}
where $p^{4}\equiv(p^{2})^{2}$, $p^{6}\equiv(p^{2})^{3}$ and $\tilde{\mathcal{C}}=\mathcal{C}-\frac{m^{2}}{2p^{2}}\psi(1)$. The above expression leads to the inclusion of following non-local term in the action (in momentum space representation)
\begin{equation}
\begin{split}
\kappa^{2}A^{\mu}[\Box\partial_{\mu}\partial_{\nu} & -\eta_{\mu\nu}\Box^{2}]\frac{1}{(4\pi)^{2}}\Bigg[\tilde{\mathcal{C}}-\left(\frac{1}{2}-\frac{3m^{2}}{2\Box}+\frac{3m^{4}}{2\Box^{2}}-\frac{m^{6}}{2\Box^{3}}\right)\ln\left(\frac{-\Box+m^{2}}{4\pi\Lambda^{2}}\right)\\
 & -\frac{3m^{2}}{2\Box}\left(\ln\left(\frac{m^{2}}{4\pi\Lambda^{2}}\right)+1\right)+\frac{3m^{4}}{2\Box^{2}}\left(\ln\left(\frac{m^{2}}{4\pi\Lambda^{2}}\right)+\frac{1}{3}\right)-\frac{1}{2}\frac{m^{6}}{\Box^{3}}\ln\left(\frac{m^{2}}{4\pi\Lambda^{2}}\right)\Bigg]A^{\nu}. 
\end{split}
\end{equation}
However, choosing $\partial_{\mu}A^{\mu}=0$ gauge leads to the following contribution 
\begin{equation}
\begin{split}
-\kappa^{2}A_{\mu}\Box^{2} & \frac{1}{(4\pi)^{2}}\Bigg[\tilde{\mathcal{C}}-\left(\frac{1}{2}-\frac{3m^{2}}{2\Box}+\frac{3m^{4}}{2\Box^{2}}-\frac{m^{6}}{2\Box^{3}}\right)\ln\left(\frac{-\Box+m^{2}}{4\pi\Lambda^{2}}\right)\\
 & -\frac{3m^{2}}{2\Box}\left(\ln\left(\frac{m^{2}}{4\pi\Lambda^{2}}\right)+1\right)+\frac{3m^{4}}{2\Box^{2}}\left(\ln\left(\frac{m^{2}}{4\pi\Lambda^{2}}\right)+\frac{1}{3}\right)-\frac{1}{2}\frac{m^{6}}{\Box^{3}}\ln\left(\frac{m^{2}}{4\pi\Lambda^{2}}\right)\Bigg]A^{\mu}.
\end{split}
\end{equation}
Taking into account this quantum correction gives the following on-shell condition
\begin{equation}
\begin{split}
p^{2}+\kappa^{2}p^{4}\frac{1}{(4\pi)^{2}} & \Bigg[\tilde{\mathcal{C}}-\left(\frac{1}{2}+\frac{3m^{2}}{2p^{2}}+\frac{3m^{4}}{2p^{4}}+\frac{m^{6}}{2p^{6}}\right)\ln\left(\frac{p^{2}+m^{2}}{4\pi\Lambda^{2}}\right)\\
+\frac{3m^{2}}{2p^{2}} & \left(\ln\left(\frac{m^{2}}{4\pi\Lambda^{2}}\right)+1\right)+\frac{3m^{4}}{2p^{4}}\left(\ln\left(\frac{m^{2}}{4\pi\Lambda^{2}}\right)+\frac{1}{3}\right)+\frac{1}{2}\frac{m^{6}}{p^{6}}\ln\left(\frac{m^{2}}{4\pi\Lambda^{2}}\right)\Bigg]=0\\
\implies p^{2}+\kappa^{2}p^{4}\frac{1}{(4\pi)^{2}} & \Bigg[\tilde{\mathcal{C}}-\left(\frac{3m^{2}}{2p^{2}}+\frac{3m^{4}}{2p^{4}}+\frac{m^{6}}{2p^{6}}\right)\ln\left(\frac{p^{2}+m^{2}}{m^{2}}\right)-\frac{1}{2}\ln\left(\frac{p^{2}+m^{2}}{4\pi\Lambda^{2}}\right)+\frac{3m^{2}}{2p^{2}}+\frac{m^{4}}{2p^{4}}\Bigg]=0.
\end{split}
\end{equation}
Although, we do not give an analytic expression of the dispersion relation, however, it can be easily checked that apart from tree-level dispersion $p^{2}=0$ there exists another dispersion relation for photons, shifted in the complex plane to a point which depends on the ratio of $\kappa,m^{2}$ to $\Lambda^{2}$ (see examples of higher derivative gauge invariant theories in \cite{thibes2017reduced, cambiaso2012massive} containing massive spin-1 particles).

The above expression also suggests that after considering one-loop correction, the propagator of photons would be $\frac{1}{p^{2}+\Sigma(p^{2})}$, where $\Sigma(p^{2})$ is the self-energy of the photon. A nice property of the theory, apart from IR modification, is that in the UV limit the propagator essentially becomes $\frac{1}{\Sigma(p^{2})}\sim\frac{1}{\kappa^{2}p^{4}\ln\left(\frac{p^{2}+m^{2}}{4\pi\Lambda^{2}}\right)}$. This shows that the degree of superficial divergence of any Feynman diagram is reduced by $(2\times\text{number of photon propagators in that diagram}$). Hence, taking into account quantum corrections at the one-loop level modifies both the UV and IR limit of the theory significantly.

\subsection{Effective interaction Vertex}
\begin{figure}
\begin{center}
\includegraphics[height=3cm, width=5cm]{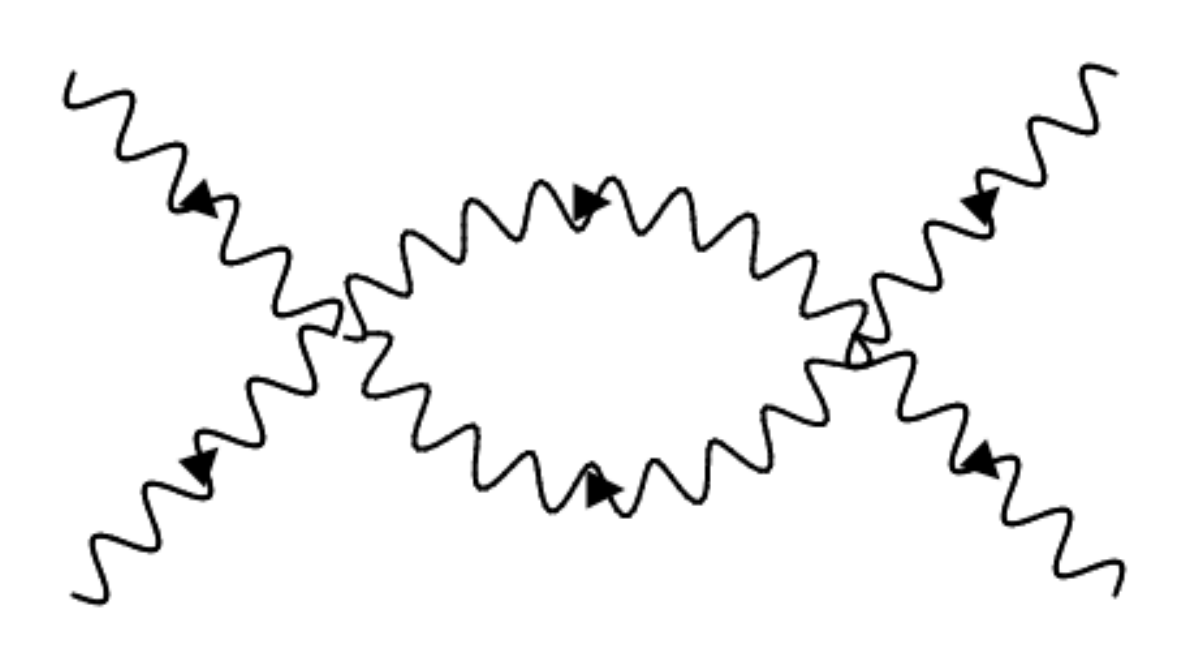}
\end{center}
\caption{1-loop vertex diagram}
\end{figure}
Figure 2 depicts the one-loop vertex diagram, which is computed next. The diagram corresponds to
\begin{equation}
-\int\frac{d^{4}k}{(2\pi)^{4}}\frac{\eta^{\alpha\gamma}}{k^{2}}\frac{\eta^{\beta\delta}}{(p-k)^{2}}\mathcal{V}_{\mu\nu\alpha\beta}(p_{1},p_{2},k,p-k)\mathcal{V}_{\gamma\delta\rho\sigma}(k,p-k,p_{3},p_{4}),
\end{equation}
where $p=p_{1}+p_{2}=p_{3}+p_{4}$. Therefore, we need to compute the following expression
\begin{equation}
\begin{split}
\eta^{\alpha\gamma}\eta^{\beta\delta} & \mathcal{V}_{\mu\nu\alpha\beta}(p_{1},p_{2},k,p-k)\mathcal{V}_{\gamma\delta\rho\sigma}(k,p-k,p_{3},p_{4})\\
 & =\Big[p_{1}.p_{2}k.(p-k)\eta_{\mu\nu}\eta^{\gamma\delta}-p_{1}.p_{2}\eta_{\mu\nu}k^{\delta}(p-k)^{\gamma}-k.(p-k)\eta^{\gamma\delta}p_{1\nu}p_{2\mu}\\
 & +p_{1\nu}p_{2\mu}k^{\delta}(p-k)^{\gamma}-p_{1}.p_{2}k.(p-k)\delta_{\mu}^{\gamma}\delta_{\nu}^{\delta}+p_{1}.p_{2}\delta_{\mu}^{\gamma}k^{\delta}(p-k)_{\nu}+p_{1}.p_{2}\delta_{\nu}^{\delta}k_{\mu}(p-k)^{\gamma}\\
 & -p_{1}.p_{2}\eta^{\gamma\delta}k_{\mu}(p-k)_{\nu}+k.(p-k)\delta_{\mu}^{\gamma}p_{1\nu}p_{2}^{\delta}-p_{2}.(p-k)\delta_{\mu}^{\gamma}p_{1\nu}k^{\delta}-p_{1\nu}p_{2}^{\delta}k_{\mu}(p-k)^{\gamma}\\
 & +p_{2}.(p-k)\eta^{\gamma\delta}p_{1\nu}k_{\mu}+k.(p-k)\delta_{\nu}^{\delta}p_{1}^{\gamma}p_{2\mu}-p_{1}^{\gamma}p_{2\mu}k^{\delta}(p-k)_{\nu}-p_{1}.k\delta_{\nu}^{\delta}p_{2\mu}(p-k)^{\gamma}\\
 & +p_{1}.k\eta^{\gamma\delta}p_{2\mu}(p-k)_{\nu}-k.(p-k)\eta_{\mu\nu}p_{1}^{\gamma}p_{2}^{\delta}+p_{2}.(p-k)\eta_{\mu\nu}p_{1}^{\gamma}k^{\delta}+p_{1}.k\eta_{\mu\nu}p_{2}^{\delta}(p-k)^{\gamma}\\
 & -\eta_{\mu\nu}\eta^{\gamma\delta}p_{1}.kp_{2}.(p-k)\Big]\frac{\kappa^{2}}{p^{2}}\\
\times\frac{\kappa^{2}}{p^{2}} & \Big[k.(p-k)p_{3}.p_{4}\eta_{\gamma\delta}\eta_{\rho\sigma}-k.(p-k)\eta_{\gamma\delta}p_{3\sigma}p_{4\rho}-p_{3}.p_{4}\eta_{\rho\sigma}k_{\delta}(p-k)_{\gamma}+k_{\delta}(p-k)_{\gamma}p_{3\sigma}p_{4\rho}\\
 & -k.(p-k)p_{3}.p_{4}\eta_{\gamma\rho}\eta_{\delta\sigma}+k.(p-k)\eta_{\gamma\rho}p_{3\sigma}p_{4\delta}+k.(p-k)\eta_{\delta\sigma}p_{3\gamma}p_{4\rho}-k.(p-k)\eta_{\rho\sigma}p_{3\gamma}p_{4\delta}\\
 & +p_{3}.p_{4}\eta_{\gamma\rho}k_{\delta}(p-k)_{\sigma}-(p-k).p_{4}\eta_{\gamma\rho}k_{\delta}p_{3\sigma}-k_{\delta}(p-k)_{\sigma}p_{3\gamma}p_{4\rho}+(p-k).p_{4}\eta_{\rho\sigma}k_{\delta}p_{3\gamma}\\
 & +p_{3}.p_{4}\eta_{\delta\sigma}k_{\rho}(p-k)_{\gamma}-k_{\rho}(p-k)_{\gamma}p_{3\sigma}p_{4\delta}-k.p_{3}\eta_{\delta\sigma}(p-k)_{\gamma}p_{4\rho}+k.p_{3}\eta_{\rho\sigma}(p-k)_{\gamma}p_{4\delta}\\
 & -p_{3}.p_{4}\eta_{\gamma\delta}k_{\rho}(p-k)_{\sigma}+(p-k).p_{4}\eta_{\gamma\delta}k_{\rho}p_{3\sigma}+k.p_{3}\eta_{\gamma\delta}(p-k)_{\sigma}p_{4\rho}-\eta_{\gamma\delta}\eta_{\rho\sigma}k.p_{3}(p-k).p_{4}\Big].
\end{split}
\end{equation}
The computation of each integral coming from such a large expression would be very cumbersome. Therefore, we compute the integral of the most general term from which, in principle, all the above terms could be computed by contractions. The most general form of the integral that we need to compute is of the following form
\begin{equation}
\begin{split}
\int\frac{d^{D}k}{(2\pi)^{D}} & \frac{1}{k^{2}(p-k)^{2}}k^{a}k^{b}(p-k)^{c}(p-k)^{d}\\
=p^{a}p^{b}p^{c}p^{d} & \int\frac{d^{D}k}{(2\pi)^{D}}\int_{0}^{1}\frac{x^{2}(1-x)^{2}}{[k^{2}+p^{2}x(1-x)]^{2}}dx \ +p^{c}p^{d}\int\frac{d^{D}k}{(2\pi)^{D}}\int_{0}^{1}dx\frac{(1-x)^{2}k^{a}k^{b}}{[k^{2}+p^{2}x(1-x)]^{2}}\\
-\int\frac{d^{D}k}{(2\pi)^{D}} & \int_{0}^{1}dx\frac{x(1-x)(k^{a}p^{b}+k^{b}p^{a})(k^{c}p^{d}+p^{c}k^{d})}{[k^{2}+p^{2}x(1-x)]^{2}}+\int\frac{d^{D}k}{(2\pi)^{D}}\int_{0}^{1}dx\frac{k^{a}k^{b}k^{c}k^{d}}{[k^{2}+p^{2}x(1-x)]^{2}}\\
 & +p^{a}p^{b}\int\frac{d^{D}k}{(2\pi)^{D}}\int_{0}^{1}dx\frac{x^{2}k^{c}k^{d}}{[k^{2}+p^{2}x(1-x)]^{2}},
\end{split}
\end{equation}
where $D=4-\varepsilon$ in dimensional regularization. Computing each integral separately and taking only finite pieces, it can be shown that above expression takes following form
\begin{equation}
[p^{a}p^{b}p^{c}p^{d}\mathcal{I}_{1}^{(\text{finite})}+(p^{c}p^{d}\eta^{ab}+p^{a}p^{b}\eta^{cd})\mathcal{I}_{2}^{(\text{finite})}+\mathcal{I}_{4}^{(\text{finite})}-\mathcal{I}_{3}^{(\text{finite})}],
\end{equation}
where
\begin{equation}
\begin{split}
\mathcal{I}_{1}^{(\text{finite})} & =\frac{1}{(4\pi)^{2}}\Big[\frac{\psi(0)}{30}-\frac{1}{30}\ln\frac{p^{2}}{4\pi\mu^{2}}+\frac{47}{900}\Big], \ \mathcal{I}_{2}^{(\text{finite})}=-\frac{\eta^{ab}p^{2}}{80(4\pi)^{2}}\left(1-\frac{1}{10}\ln\frac{p^{2}}{4\pi\mu^{2}}+\frac{13}{75}\right)\\
\mathcal{I}_{3}^{(\text{finite})} & =-\frac{\eta^{ac}p^{b}p^{d}+\eta^{ad}p^{b}p^{c}+\eta^{bc}p^{a}p^{d}+\eta^{bd}p^{a}p^{c}}{4}\frac{p^{2}}{(4\pi)^{2}}\Big[\frac{1}{30}-\frac{1}{15}\ln\frac{p^{2}}{4\pi\mu^{2}}+\frac{47}{450}\Big]\\
\mathcal{I}_{4}^{(\text{finite})} & =-\frac{1}{\alpha}(\eta^{ab}\eta^{cd}+\eta^{ac}\eta^{bd}+\eta^{ad}\eta^{bc})\frac{(p^{2})^{2}}{(4\pi)^{2}}\left(\frac{\psi(0)}{30}-\frac{1}{15}\ln\frac{p^{2}}{4\pi\mu^{2}}+\frac{47}{900}\right)\\
 & -\frac{2p^{4}}{\alpha(4\pi)^{2}}(\eta^{ab}\eta^{cd}+\eta^{ac}\eta^{bd}+\eta^{ad}\eta^{bc})\left(\frac{1}{30}-\frac{1}{15}\ln\frac{p^{2}}{4\pi\mu^{2}}-\frac{47}{450}\right),  
\end{split}
\end{equation}
and $\alpha$ is just a numerical coefficient. In the effective action, in place of $p^{2}$, one just needs to write $-\Box$ and each $p^{a}$ term must be replaced by $-i\partial^{a}$. This shows that the quantum correction of the 4-point vertex in the effective action also contains non-local terms because of the presence of $\ln\frac{-\Box}{4\pi\mu^{2}}$, where $\mu$ is the momentum scale up to which this theory is valid in a perturbative manner.

\section{Conclusion}
The principal aim of the present article has been to bring out the various features
of the interaction between photons and gravitons that can be used in astrophysical
observations. The effective action for photons, developed here, captures possible interactions between photons and gravitons at the quantum level. Furthermore, it is shown that through the weak measurement protocol one can enhance the strength of the scattering amplitudes or cross-sections of the scattering process between multiple photons which would make it possible to be measured in the laboratory. Polarization measurement of photons will also be able to capture this interaction, an idea also suggested in \cite{Ghosh_2016, hansen2015search}. Though S-matrix elements of photon-graviton interaction were calculated before in \cite{weinberg1964photons, weinberg1965photons, papini1977gravitons}, here we have used a different approach in which instead of measuring gravitons directly, inferences can be drawn from measurements on photon states, a task comparatively easier to achieve in current experimental scenarios.

We have also shown how Maxwell's equations get modified in the presence of gravity. This can capture the properties of the source of GW, such as compact objects, binary mergers, in terms of their stress-energy tensors. Vacuum birefringence property \cite{hattori2013vacuum, ataman2018vacuum, nakamiya2017probing, Kruglov_2015} is seen to have emerged from the modified Maxwell's equations where polarizability and magnetization non-linearly depend on the electric and magnetic fields. The results due to gravitons from GR were consistently compared with massive gravity theory which is an IR modified version of GR. This will put a bound on the mass of gravitons, and by studying graviton-photon interaction using photon's effective action in (\ref{eff2}), one can suitably modify the IR domain of GR. 

Finally, the modified dispersion of photons has been computed by taking into account one-loop quantum corrections both in the case of massive and massless gravitons. This dispersion is shown to be scale-dependent (these scales are basically Planck length scale or Planck mass and graviton mass scale), coming from the presence of non-local terms in the quantum effective action at the quadratic level. It is also shown in (\ref{dispersion0}) that the light-matter coupling gives a non-trivial dispersion of photons that depends on the details of the stress-energy tensor of GW sources.

The backreaction, the effect of the stress-energy of quantum fields on the curvature of the background spacetime on which fields are propagating is described by the semiclassical Einstein equation $G_{\mu\nu}=\langle:\hat{T}_{\mu\nu}:\rangle$ \cite{hack2010backreaction} where $: \ :$ is the normal-ordering operation. In the present discussion, we do not consider the backreaction since we restricted our discussion to the low energy theory. This follows from the fact that the higher-order interaction terms between gravitons and photons are omitted from the action as the mass-dimension of the gauge fields is one. In our construction of this low energy theory, the relevant degrees of freedom of the system are identified along with the interactions that are compatible with the expected symmetries. However, at low energies, the most important terms in the action are those that are least suppressed by powers of the scale $\kappa$.

\section{Acknowledgement}
SM would like to thank IISER Kolkata for supporting this work through a doctoral fellowship and also wants to thank IIT-Jodhpur for their kind hospitality during his stay in their campus where a part of the work and discussion regarding the work were done. ABS is grateful to Ministry of Human Resources Development, Govt. of India, for research fellowship through Center of Excellence in Space Sciences India.

\bibliographystyle{unsrt}
\bibliography{bibtexfile}

\section{Appendix}\label{appendix}
\subsection{Free Fierz-Pauli Action}
The Fierz-Pauli action that describes a massive spin-2 particle in flat spacetime by a symmetric rank-2 tensor is 
\begin{equation}\label{Fierz-Pauli}
S=\int d^{D}x\Big[-\frac{1}{2}\partial_{\lambda}h_{\mu\nu}\partial^{\lambda}h^{\mu\nu}
+\partial_{\mu}h_{\nu\lambda}\partial^{\nu}h^{\mu\lambda}-\partial_{\mu}h^{\mu\nu}\partial_{\nu}h+\frac{1}{2}\partial_{\lambda}h\partial^{\lambda}h-\frac{1}{2}m^{2}(h_{\mu\nu}h^{\mu\nu}-h^{2})\Big].
\end{equation}
Note that when $m=0$ it becomes the linearized Einstein-Hilbert action which is invariant under the following gauge transformation
\begin{equation}\label{A.0}
\delta h_{\mu\nu}=\partial_{\mu}\xi_{\nu}+\partial_{\nu}\xi_{\mu}.
\end{equation}
Though the above action is not gauge invariant, using Stueckelberg's trick, it can be made gauge invariant.

\subsection{Equations of motion and Degrees of freedom}
The equations of motion corresponding to the action (\ref{Fierz-Pauli}) is 
\begin{equation}\label{A.1}
\Box h_{\mu\nu}-\partial_{\lambda}\partial_{\mu}h_{ \ \nu}^{\lambda}-\partial_{\lambda}\partial_{\nu}h_{ \ \mu}^{\lambda}+\eta_{\mu\nu}\partial_{\lambda}\partial_{\sigma}h^{\lambda\sigma}+\partial_{\mu}\partial_{\nu}h-\eta_{\mu\nu}\Box h=m^{2}(h_{\mu\nu}-\eta_{\mu\nu}h).
\end{equation}
The \textit{l.h.s} of the above equation consists of the linearized form of the Einstein tensor $G_{\mu\nu}=\mathcal{R}_{\mu\nu}-\frac{1}{2}g_{\mu\nu}\mathcal{R}$ and has zero-divergence. Hence, acting $\partial^{\mu}$ on it, we obtain
\begin{equation}
m^{2}(\partial^{\mu}h_{\mu\nu}-\partial_{\nu}h)=0.
\end{equation}
Since $m\neq0$, we obtain the following constraints
\begin{equation}\label{A.2}
\partial^{\mu}h_{\mu\nu}=\partial_{\nu}h.
\end{equation}
Plugging (\ref{A.2}) into (\ref{A.1}) gives the following equation 
\begin{equation}
\Box h_{\mu\nu}-\partial_{\mu}\partial_{\nu}h=m^{2}(h_{\mu\nu}-\eta_{\mu\nu}h).
\end{equation}
Trace of the above equation gives 
\begin{equation}
\Box h-\Box h=-3m^{2}h=0\implies h=0,
\end{equation}
which means $h_{\mu\nu}$ is traceless and transverse. Further, using traceless and transverse property of $h_{\mu\nu}$, we obtain the following equation of motion
\begin{equation}
(\Box-m^{2})h_{\mu\nu}=0.
\end{equation}
Therefore, the equations of motion give us ten wave equations and five constraints. Hence, in four dimensions, we have five degrees of freedom; these degrees of freedom are nothing but the massive spin-2 gravitons.

\subsection{Propagator}
In order to find the propagator of massive gravitons, we express the Fierz-Pauli action in the following form
\begin{equation}
S=\int d^{4}x \frac{1}{2}h_{\mu\nu}\mathcal{O}^{\mu\nu\alpha\beta}h_{\alpha\beta},
\end{equation}
where
\begin{equation}
\mathcal{O}_{ \ \ \alpha\beta}^{\mu\nu}=(\eta_{ \ \alpha}^{(\mu}\eta_{ \ \beta}^{\nu)}-\eta^{\mu\nu}\eta_{\alpha\beta})(\Box-m^{2})-2\partial^{(\mu}\partial_{(\alpha}\eta_{ \ \beta)}^{\nu)}+\partial^{\mu}\partial^{\nu}\eta_{\alpha\beta}+\partial_{\alpha}\partial_{\beta}\eta^{\mu\nu}.
\end{equation} 
Therefore, the propagator denoted by $\mathcal{D}_{\alpha\beta,\sigma\lambda}$ is defined in the following way
\begin{equation}
\mathcal{O}^{\mu\nu,\alpha\beta}\mathcal{D}_{\alpha\beta,\sigma\lambda}=\frac{i}{2}(\delta_{\sigma}^{\mu}\delta_{\lambda}^{\nu}+\delta_{\sigma}^{\nu}\delta_{\lambda}^{\mu}).
\end{equation}
As a result, in momentum space the propagator takes the following form
\begin{equation}
\mathcal{D}_{\alpha\beta,\sigma\lambda}=-\frac{i}{p^{2}+m^{2}}\Big[\frac{1}{2}(\mathcal{P}_{\alpha\sigma}\mathcal{P}_{\beta\lambda}+\mathcal{P}_{\alpha\lambda}\mathcal{P}_{\beta\sigma})-\frac{1}{3}\mathcal{P}_{\alpha\beta}\mathcal{P}_{\sigma\lambda}\Big],
\end{equation}
where
\begin{equation}
\mathcal{P}_{\alpha\beta}=\eta_{\alpha\beta}+\frac{p_{\alpha}p_{\beta}}{m^{2}}.
\end{equation}
It shows that in high energy limit (large momenta limit), the graviton propagator behaves as 
\begin{equation}
\mathcal{D}_{\alpha\beta,\sigma\lambda}\simeq\frac{1}{p^{2}+m^{2}}\frac{p_{\alpha}p_{\beta}p_{\sigma}p_{\lambda}}{m^{4}}\simeq\frac{p^{2}}{m^{4}},
\end{equation}
which implies that standard power counting rules are no longer valid. This seems to suggest that the renormalizability of this theory is not guaranteed. However, it is not true which can be shown explicitly using Stueckelberg's trick.

\subsection{Stueckelberg's trick}
Here we briefly review the technique introduced by Stueckelberg to the massive gravity action with a source in order to restore the gauge symmetry. For the sake of simplicity, we write massless terms in the Lagrangian density separately
\begin{equation}\label{A.3}
S=\int d^{4}x\Big[\mathcal{L}_{m=0}-\frac{1}{2}m^{2}(h_{\mu\nu}h^{\mu\nu}-h^{2})+\kappa h_{\mu\nu}T^{\mu\nu}\Big].
\end{equation}
It is well-known that the massless gravitons have gauge symmetry which is broken due to the presence of a mass term in the above action. Now we introduce a new field $V_{\mu}$, known as the Stueckelberg field using the following field redefinition
\begin{equation}
h_{\mu\nu}\rightarrow h_{\mu\nu}+\partial_{\mu}V_{\nu}+\partial_{\nu}V_{\mu}.
\end{equation}
Note that under this field redefinition, $\mathcal{L}_{m=0}$ remains invariant as it is invariant under the infinitesimal diffeomorphism (\ref{A.0}), however, the other terms do change and we obtain
\begin{equation}\label{A.4}
\begin{split}
S & =\int d^{4}x\Bigg[\mathcal{L}_{m=0}-\frac{1}{2}m^{2}\Big[(h_{\mu\nu}+\partial_{\mu}V_{\nu}+\partial_{\nu}V_{\mu})(h^{\mu\nu}+\partial^{\mu}V^{\nu}+\partial^{\nu}V^{\mu})-(h+2\partial_{\mu}V^{\mu})^{2}\Big]+\kappa h_{\mu\nu}T^{\mu\nu}-2\kappa V_{\mu}\partial_{\nu}T^{\mu\nu}\Bigg]\\
 & =\int d^{4}x\Big[\mathcal{L}_{m=0}-\frac{1}{2}m^{2}(h_{\mu\nu}h^{\mu\nu}-h^{2})-2m^{2}(h_{\mu\nu}\partial^{\mu}V^{\nu}-h\partial_{\mu}V^{\mu})-\frac{1}{2}m^{2}\bar{F}_{\mu\nu}\bar{F}^{\mu\nu}+\kappa h_{\mu\nu}T^{\mu\nu}-2\kappa V_{\mu}\partial_{\nu}T^{\mu\nu}\Big],
\end{split}
\end{equation}
where 
\begin{equation}
\bar{F}_{\mu\nu}\equiv\partial_{\mu}V_{\nu}-\partial_{\nu}V_{\mu}.
\end{equation}
Note that the above action (\ref{A.4}) has the following gauge symmetry
\begin{equation}
h_{\mu\nu}\rightarrow h_{\mu\nu}+\partial_{\mu}\xi_{\nu}+\partial_{\nu}\xi_{\mu}, \ \ V_{\mu}\rightarrow V_{\mu}-\xi_{\mu}.
\end{equation}
We can fix it to $V_{\mu}=0$ and recover the original action. Therefore, both the actions (\ref{A.3}) and (\ref{A.4}) are equivalent. It is important to note here that if we try to take the $m\rightarrow0$ limit, it does not go smoothly as one degree of freedom is lost. Hence, we need to do a similar kind of transformation one more time. This is of the following form
\begin{equation}
V_{\mu}\rightarrow V_{\mu}+\partial_{\mu}\phi.
\end{equation}
With the above transformation, the previous action (\ref{A.4}) becomes
\begin{equation}\label{A.5}
\begin{split}
S & =\int d^{4}x\Big[\mathcal{L}_{m=0}-\frac{1}{2}m^{2}(h_{\mu\nu}h^{\mu\nu}-h^{2})-\frac{1}{2}m^{2}\bar{F}_{\mu\nu}\bar{F}^{\mu\nu}-2m^{2}(h_{\mu\nu}\partial^{\mu}V^{\nu}-h\partial_{\mu}V^{\mu})\\
 & -2m^{2}(h_{\mu\nu}\partial^{\mu}\partial^{\nu}\phi-h\Box\phi)+\kappa h_{\mu\nu}T^{\mu\nu}-2\kappa V_{\mu}\partial_{\nu}T^{\mu\nu}+2\kappa\phi\partial_{\mu}\partial_{\nu}T^{\mu\nu}\Big].
\end{split}
\end{equation}
The resultant action has two independent gauge symmetries
\begin{equation}
\begin{split}
h_{\mu\nu}\rightarrow h_{\mu\nu}+\partial_{\mu}\xi_{\nu}+\partial_{\nu}\xi_{\mu}, \ & \ \ V_{\mu}\rightarrow V_{\mu}-\xi_{\mu}\\
V_{\mu}\rightarrow V_{\mu}+\partial_{\mu}\Lambda, & \ \ \phi\rightarrow\phi-\Lambda.
\end{split}
\end{equation}
As before, we can fix the gauge $\phi=0$ and recover back the action (\ref{A.4}) which implies that the action (\ref{A.5}) is equivalent to the action (\ref{A.4}). Hence, with the new additional fields and gauge symmetries, the new action does the same job as the original one in (\ref{A.3}).

Using the following set of scalings
\begin{equation}
V_{\mu}\rightarrow\frac{1}{m}V_{\mu}, \ \phi\rightarrow\frac{\phi}{m^{2}},
\end{equation}
the action (\ref{A.5}) can be expressed as
\begin{equation}\label{A.5-6}
\begin{split}
S & =\int d^{4}x\Big[\mathcal{L}_{m=0}-\frac{1}{2}m^{2}(h_{\mu\nu}h^{\mu\nu}-h^{2})-\frac{1}{2}\bar{F}_{\mu\nu}\bar{F}^{\mu\nu}-2m(h_{\mu\nu}\partial^{\mu}V^{\nu}-h\partial_{\mu}V^{\mu})\\
 & -2(h_{\mu\nu}\partial^{\mu}\partial^{\nu}\phi-h\Box\phi)+\kappa h_{\mu\nu}T^{\mu\nu}-2\frac{\kappa}{m} V_{\mu}\partial_{\nu}T^{\mu\nu}+2\frac{\kappa}{m^{2}}\phi\partial_{\mu}\partial_{\nu}T^{\mu\nu}\Big],
\end{split}
\end{equation}
with the following gauge symmetries
\begin{equation}
\begin{split}
h_{\mu\nu}\rightarrow h_{\mu\nu}+\partial_{\mu}\xi_{\nu}+\partial_{\nu}\xi_{\mu}, \ & \ \ V_{\mu}\rightarrow V_{\mu}-m\xi_{\mu}\\
V_{\mu}\rightarrow V_{\mu}+m\partial_{\mu}\Lambda, & \ \ \phi\rightarrow\phi-m^{2}\Lambda.
\end{split}
\end{equation}
Since the stress-energy tensor of the source is conserved $\partial_{\mu}T^{\mu\nu}=0$, in $m\rightarrow 0$ limit, we obtain the following expression for the resulting action
\begin{equation}\label{A.6}
S=\int d^{4}x\Big[\mathcal{L}_{m=0}-\frac{1}{2}\bar{F}_{\mu\nu}\bar{F}^{\mu\nu}-2(h_{\mu\nu}\partial^{\mu}\partial^{\nu}\phi-h\Box\phi)+\kappa h_{\mu\nu}T^{\mu\nu}\Big].
\end{equation}
In order to count the total number of degrees of freedom, we make a conformal transformation which is of the following form
\begin{equation}\label{conformal trans}
\begin{split}
(\eta_{\mu\nu}+h_{\mu\nu}) & =\Omega(\eta_{\mu\nu}+h'_{\mu\nu})=(1+\Pi)(\eta_{\mu\nu}+h'_{\mu\nu})\\
 & =\eta_{\mu\nu}+h'_{\mu\nu}+\Pi\eta_{\mu\nu}\\
\implies h_{\mu\nu} & =h'_{\mu\nu}+\Pi\eta_{\mu\nu},
\end{split}
\end{equation}
where $\Pi$ is another scalar. Hence, this is nothing but the redefinition of the field $h_{\mu\nu}$. Under this transformation, the massless Lagrangian $\mathcal{L}_{m=0}$ becomes
\begin{equation}
\begin{split}
\mathcal{L}_{m=0}[h] & =\mathcal{L}_{m=0}[h']-\partial_{\lambda}\Pi\partial^{\lambda}h'-2\partial_{\lambda}\Pi\partial^{\lambda}\Pi+2\partial_{\mu}\Pi\partial_{\lambda}h'^{\mu\lambda}+\partial_{\mu}\Pi\partial^{\mu}\Pi-\partial_{\mu}\Pi\partial^{\mu}h'\\
 & -4\partial_{\mu}h'^{\mu\nu}\partial_{\nu}\Pi-4\partial_{\mu}\Pi\partial^{\mu}\Pi+4\partial_{\lambda}\Pi\partial^{\lambda}h'+8\partial_{\lambda}\Pi\partial^{\lambda}\Pi+\kappa\Pi T\\
 & =\mathcal{L}_{m=0}[h']+2\Big[\partial_{\mu}\partial^{\mu}h'-\partial_{\nu}h'^{\mu\nu}\partial_{\mu}\Pi+\frac{3}{2}\partial_{\mu}\Pi\partial^{\mu}\Pi\Big]+\kappa\Pi T,
\end{split}
\end{equation}
whereas the action in (\ref{A.6}) becomes
\begin{equation}
\begin{split}
S & =\int d^{4}x\Bigg[\mathcal{L}_{m=0}[h']+2\Big[\partial_{\mu}\partial^{\mu}h'-\partial_{\nu}h'^{\mu\nu}\partial_{\mu}\Pi+\frac{3}{2}\partial_{\mu}\Pi\partial^{\mu}\Pi\Big]\\
 & -\frac{1}{2}\bar{F}_{\mu\nu}\bar{F}^{\mu\nu}-2(h'_{\mu\nu}\partial^{\mu}\partial^{\nu}\phi-h'\Box\phi)+6\Pi\Box\phi+\kappa h'_{\mu\nu}T^{\mu\nu}+\kappa\Pi T\Bigg].
\end{split}
\end{equation}
Further, considering $\Pi=\phi$ cancels all the coupled tensor-scalar terms. An  integration by parts, yields the following action
\begin{equation}
\begin{split}
S & =\int d^{4}x\Big[\mathcal{L}_{m=0}[h']-\frac{1}{4}\bar{F}_{\mu\nu}\bar{F}^{\mu\nu}-\frac{1}{2}\partial_{\mu}\phi'\partial^{\mu}\phi+\kappa h'_{\mu\nu}T^{\mu\nu}+\frac{1}{\sqrt{6}}\phi' T\Big],
\end{split}
\end{equation}
with the following two independent gauge symmetries
\begin{equation}
\begin{split}
h'_{\mu\nu} & \rightarrow h'_{\mu\nu}+\partial_{\mu}\xi_{\nu}+\partial_{\nu}\xi_{\mu}\\
V'_{\mu} & \rightarrow V'_{\mu}+\partial_{\mu}\Lambda,
\end{split}
\end{equation}
where $V_{\mu}\rightarrow V'_{\mu}=\sqrt{2}V_{\mu}, \ \phi\rightarrow\phi'=\sqrt{\frac{3}{2}}\phi$.

Hence, in four dimensions, we have one massless graviton which possesses two degrees of freedom, one massless vector field which also possesses two degrees of freedom, and one massless scalar, in total making 5 degrees of freedom.

If we now consider the massive action (\ref{A.5-6}) and make the same transformation (\ref{conformal trans}), we obtain the following action
\begin{equation}
\begin{split}
S & =\int d^{4}x\Big[\mathcal{L}_{m=0}[h']-\frac{1}{2}m^{2}(h'_{\mu\nu}h'^{\mu\nu}-h'^{2})-\frac{1}{2}\bar{F}_{\mu\nu}\bar{F}^{\mu\nu}+3\phi(\Box+2m^{2})\phi-2m(h'_{\mu\nu}\partial^{\mu}V^{\nu}-h'\partial_{\mu}V^{\mu})\\
 & +3(2m\phi\partial_{\mu}V^{\mu}+m^{2}h'\phi)+\kappa h'_{\mu\nu}T^{\mu\nu}+\kappa\phi T-\frac{2}{m}\kappa V_{\mu}\partial_{\nu}T^{\mu\nu}+\frac{2}{m^{2}}\kappa\phi\partial_{\mu}\partial_{\nu}T^{\mu\nu}\Big].
\end{split}
\end{equation}
The gauge symmetries now read
\begin{equation}
\begin{split}
\delta h'_{\mu\nu}=\partial_{\mu}\xi_{\nu}+\partial_{\nu}\xi_{\mu}+m\Lambda\eta_{\mu\nu}, & \ \ \delta V_{\mu}=-m\xi_{\mu}+\partial_{\mu}\Lambda\\
\delta V_{\mu}=\partial_{\mu}\Lambda, & \ \ \delta\phi=m\Lambda.
\end{split}
\end{equation}
We now add two gauge fixing terms to the action 
\begin{equation}
\begin{split}
S_{GF1} & =-\int d^{4}x\left(\partial^{\nu}h'_{\mu\nu}-\frac{1}{2}\partial_{\mu}h'+mV_{\mu}\right)^{2}, \ S_{GF2}=-\int d^{4}x\left(\partial_{\mu}V^{\mu}+m\left(\frac{1}{2}h'+3\phi\right)\right)^{2}.
\end{split}
\end{equation}
Introduction of these gauge fixing terms make the action diagonalized 
\begin{equation}
\begin{split}
S+S_{GF1}+S_{GF2} & =\int d^{4}x\Big[\frac{1}{2}h'_{\mu\nu}(\Box-m^{2})h'^{\mu\nu}-\frac{1}{4}h'(\Box-m^{2})h'+V_{\mu}(\Box-m^{2})V^{\mu}+3\phi(\Box-m^{2})\phi\\
 & +\kappa h'_{\mu\nu}T^{\mu\nu}+\kappa\phi T-\frac{2}{m}\kappa V_{\mu}\partial_{\nu}T^{\mu\nu}+\frac{2}{m^{2}}\kappa\phi\partial_{\mu}\partial_{\nu}T^{\mu\nu}\Big].
\end{split}
\end{equation}
This is the Eq. (\ref{action2}) in the main text. As a consequence, the propagators of $h'_{\mu\nu}, \ V_{\mu}, \ \phi$ in the momentum space become,
\begin{equation}
\begin{split}
-\frac{i}{p^{2}+m^{2}}\frac{1}{2} & (\eta_{\mu\alpha}\eta_{\nu\beta}+\eta_{\mu\beta}\eta_{\nu\alpha}-\eta_{\mu\nu}\eta_{\alpha\beta}), \ -\frac{i}{2}\frac{\eta_{\mu\nu}}{p^{2}+m^{2}}, \ -\frac{i}{6(p^{2}+m^{2})},
\end{split}
\end{equation}
which all behave as $\frac{1}{p^{2}}$ for large momenta. Therefore, the standard power counting arguments can be used in order to renormalize this theory.

\end{document}